\documentclass{JHEP3}
\input epsf
\usepackage{epsfig}
\usepackage{amssymb}
\usepackage[active]{srcltx}

\newcommand{\bea}{\begin{eqnarray}}
\newcommand{\eea}{\end{eqnarray}}
\newcommand{\bean}{\begin{eqnarray*}}
\newcommand{\eean}{\end{eqnarray*}}

\def\W #1{\widetilde{#1}}

\def\ket#1{\left| #1\right\rangle}
\def\gb #1{ \left\langle #1 \right]}
\def\tgb #1{ \left[ #1 \right\rangle}
\def\vev#1{\left\langle #1 \right\rangle}

\def\det{\mathop{\rm det}}

\def\a{{\alpha}}

\def\la{\lambda}
\def\eps{\epsilon}

\def\ord{{\cal O}}

\def\Label#1{\label{#1}%
  \smash{\hbox to0pt{\raise1ex\hbox{\tiny[#1]}\hss}}}

\preprint{ITFA-2006-49
\\
{\tt hep-ph/0612089}}
\title{ Unitarity Cuts with Massive Propagators and Algebraic Expressions for Coefficients}
\author{Ruth Britto$^{a}$ and Bo Feng$^{b,c}$\\
~~~~\\
$^a$Institute for Theoretical Physics, University of Amsterdam \\
Valckenierstraat 65, 1018 XE Amsterdam, The Netherlands\\
$^b$Center of Mathematical Science, Zhejiang University, Hangzhou, China\\
$^c$Blackett Laboratory \& The Institute for Mathematical Sciences \\Imperial College, London, SW7 2AZ, UK\\}
\abstract{In the first part of this paper, we extend the $d$-dimensional unitarity cut method of hep-ph/0609191 to cases with massive propagators.
We present formulas for integral reduction with which one can obtain coefficients of all pentagon, box, triangle and massive bubble integrals.
In the second part of this paper, we present a detailed study of the phase space integration for unitarity cuts.  We carry out spinor integration in generality and give algebraic expressions for coefficients, intended for  automated evaluation.
}

\begin{document}
%%%%%%%%%%%%%%%%%%%%%%%%%%%%%%%%%%%%%%%%%%%%%%%%%%%%%%%%%%%%%%%%%%

%%%%%%%%%%%%%%%%%%%%%%%%%%%%%%%
\section{Introduction}
%%%%%%%%%%%%%%%%%%%%%%%%%%%%%%%

With the approach of the Large Hadron Collider experiments, accurate
descriptions of particle physics will require knowledge of
one-loop cross sections.  Computational complexity increases
dramatically with the number of legs, even at the amplitude level.
It is desirable to find a simple and fast algorithm for these
computations. There has been notable progress in the last couple of
years \cite{delAguila:2004nf,Ossola:2006us,Ellis:2006ss,Anastasiou:2004vj,
Britto:2004nc,Britto:2005ha,Britto:2006sj,Berger:2006ci,
Brandhuber:2005jw,Xiao:2006vr,Mastrolia:2006ki}.

The unitarity method introduced in~\cite{Bern:1994zx} seeks to
compute amplitudes by applying a unitarity cut to an amplitude on
one hand, and its expansion in a basis of master integrals on the
other~\cite{PV}.  With knowledge of the basis and the general
structure of the coefficients in the expansion, the coefficients can
be constrained.

The holomorphic anomaly \cite{Cachazo:2004by}
reduces the problem of phase space integration to one of algebraic manipulation, namely evaluating residues of a complex function.  By applying this operation within the unitarity method, coefficients can be extracted systematically.
In this manner, a method was introduced to evaluate any finite four-dimensional unitarity cut and systematically derive compact expressions for the coefficients~\cite{Britto:2005ha,Britto:2006sj}.

When working with four-dimensional cuts, one loses information of possible rational terms that are cut-free in four dimensions.
However, unitarity methods can find rational terms as well, if we
carry out the integral in $d=(4-2\eps)$ dimensions to higher orders in $\eps$~\cite{vanNeerven:1985xr}.  This program was developed in \cite{Bern:1995db,Bern:1996je,Bern:1996ja,Brandhuber:2005jw}.

Recently, a general $d$-dimensional unitarity cut method was developed for one-loop amplitudes~\cite{Anastasiou:2006jv,Anastasiou:2006gt}.  In this method, coefficients are extracted by first separating and performing a four-dimensional integral by a technique of choice, and then identifying the integral over the remaining $d-4$ dimensions with a coefficient using recursive dimensional shift identities.  In principle one can work out the coefficients to all orders in $\eps$.
 In fact, for cases with only massless propagators, this
method is a complete alternative to Passarino-Veltman reduction.

  In the first part of this paper (Sections 2 and 3) we generalize the work of \cite{Anastasiou:2006jv,Anastasiou:2006gt} to cases where propagators have nonzero, non-uniform masses.  We present formulas for integral reduction from which one can obtain coefficients of all scalar pentagon, box, triangle and massive bubble integrals.
Pentagons in arbitrary dimensions are truly independent master integrals, though in four dimensions they can be expressed in terms of boxes up to $\ord(\eps)$.
Amplitudes allowing massive propagators will most generally have tadpole and massless-bubble master integrals as well.  Unitarity methods cannot access this information directly, but other physical considerations pin it down.

In the second part (Section 4), we further simplify the technique of spinor integration over phase space by carrying out some intermediate steps in generality.  These steps include expressing the integrand as a total derivative and a possible Feynman-parameter integral.  We give algebraic expressions for the coefficients of the cut-containing basis integrals.  Our formulas should be suitable for programming.  To this end we include some possibly helpful identities in the appendix, as well as a discussion of the issue of multiple poles.

We close with a discussion (Section 5) of the tree-level input and possible alternative approaches.  We compare our approach of spinor integration to the procedure of Ossola, Papadopoulos and Pittau \cite{Ossola:2006us} of evaluating integrands at particular values of loop momentum and solving a system of linear equations.  Both our method and theirs algebraic substitution at the integrand level.  We claim that the complexity is equivalent.

% % % % % % % %

%%%%%%%%%%%%%%%%%%%%%%%%%%%%%%%
\section{Unitarity cuts with massive propagators}
%%%%%%%%%%%%%%%%%%%%%%%%%%%%%%%

Here we develop the program of double-cut phase
space integration in $(4-2\eps)$ dimensions \cite{Bern:1995db,Bern:1996je,Bern:1996ja,Brandhuber:2005jw,Anastasiou:2006jv,Anastasiou:2006gt} by generalizing to the case of propagators with nonzero and different masses.

One difference from the massless case is in
 the basis of master integrals. If all
propagators are massless, the basis consists of scalar pentagons, boxes, triangles
and bubbles. But if some propagators are massive, there are
tadpoles and massless bubbles (i.e., the case $K^2=0$ for bubbles).
The tadpoles are \cite{Bern:1995db}
\bea I_1 & \sim & m^{2-2\eps} { \Gamma[1+\eps]\over \eps
(\eps-1)} \\
& = & {m^2 \over \eps} + m^2 (1-\gamma-2\log(m))+\ord(\eps),
~~~~\label{One-point} \eea
while for massless bubbles we find
\bea I_2(K=0) &\sim& {M_1^{2-2\eps}-M_2^{2-2\eps} \over M_1^2 - M_2^2
}
{ \Gamma[1+\eps]\over \eps (1-\eps)}\\
&=&
{1 \over \eps} + (1-\gamma)-{2 \over M_1^2-M_2^2}
(M_1^2\log(M_1)-M_2^2\log(M_2))+\ord(\eps)
\\ & {\rm or} &
{1 \over \eps} -(\gamma  + 2 \log M_1)
+\ord(\eps)~~~~{\rm if}~~M_1=M_2.
\nonumber
\eea
It is evident from the lack of all momentum dependence that unitarity methods alone can never detect these two functions.

However, as discussed in \cite{Bern:1995db}, it is possible in some cases to
address this difficulty by considering the known divergent behavior
of the amplitude. For massless gauge theory amplitudes, the
conditions are such that quadratic divergences cancel and the remaining
(logarithmic) divergence matches the known value.  With these
conditions we can fix coefficients at leading order in $\eps$. In more general cases,  we
might need some additional information.
This could come, for example, from taking the heavy mass limit of one propagator \cite{babis}.
We shall not discuss the tadpoles and
massless bubbles further in this paper.

%%%%%%%%%%%%%%%%%%%%%%%%%%%%%
\subsection{Cuts of Scalar Integrals}
%%%%%%%%%%%%%%%%%%%%%%%%%%%%%

We define the
 $n$-point scalar function with non-uniform masses as follows:\footnote{ Note that, for ease of presentation, we are omitting the prefactor $i(-1)^{n+1}(4\pi)^{D/2}$ (which was included for example in \cite{Bern:1995db}).}
\bea   I_n(M_1,M_2,m_1,\ldots,m_{n-2})  \equiv
 \int {d^{4-2\eps} p \over
(2\pi)^{4-2\eps}}{1\over (p^2-M_1^2) ((p-K)^2-M_2^2)
\prod_{j=1}^{n-2} ((p-P_j)^2-m_{j}^2)}.~~~\label{n-scalar} \eea
This definition is written with a view towards taking the unitarity cut.  The masses of the cut propagators are $M_1$ and $M_2$, and the momentum flowing through the cut is $K$.  The other momenta $P_j$ and masses $m_j$ are the ones necessary to complete the scalar function, and not necessarily in cyclic order.

Our calculations are done in the ``Four Dimensional Helicity'' (FDH) scheme, i.e. all external momenta $K_i$ are $4$-dimensional
and only the internal momentum $p$ is $(4-2\eps)$-dimensional.
 Thus it is useful to write $p=\W \ell+\vec{\mu}$,
where $\W \ell$ is $4$-dimensional and $\vec{\mu}$ is
$(-2\eps)$-dimensional.
\bean \int {d^{4-2\eps} p\over (2\pi)^{4-2\eps}} & = & \int {d^{4}
\W\ell\over (2\pi)^{4}}\int {d^{-2\eps} \ell_\eps\over
(2\pi)^{-2\eps}} = \int {d^{4} \W\ell\over (2\pi)^4}
{(4\pi)^{\eps}\over \Gamma(-\eps)} \int  d\mu^2
(\mu^2)^{-1-\eps}.
\eean
Then the scalar function as defined in (\ref{n-scalar}) becomes
\bea   I_n(M_1,M_2,m_1,\ldots,m_{n-2})  = & &  {(4\pi)^{\eps}\over \Gamma(-\eps)}
\int d\mu^2 (\mu^2)^{-1-\eps}\int {d^{4} \W\ell\over (2\pi)^4}
 \nonumber \\
& &
{1\over (\W \ell^2-M_1^2-\mu^2) ((\W \ell-K)^2-M_2^2-\mu^2)
\prod_{j=1}^{n-2} ((\W\ell-P_j)^2-m_{j}^2-\mu^2)}.
~~~\label{n-scalar-1} \eea
Following the setup of \cite{Anastasiou:2006jv}, we decompose the 4-momentum into a null component and a component proportional to the cut momentum $K$.
\bea \W \ell= \ell+z K,~~~~~\ell^2=0,~~~\Longrightarrow \int d^4\W
\ell= \int dz ~d^4\ell ~\delta^+(\ell^2) (2 \ell \cdot K).
~~~\label{changing} \eea
We define the  ``signature of the cut,'' $\Delta[K,M_1,M_2]$, as
\bea
 { \Delta[K,M_1,M_2] } \equiv
{ (K^2)^2+(M_1^2)^2+(M_2^2)^2-2 K^2 M_1^2 -2 K^2
M_2^2- 2M_1^2 M_2^2 },~~~~\label{delta-cut}\eea
and a dimensionless parameter $u$ to be
\bea u\equiv {4 K^2\mu^2 \over \Delta[K,M_1,M_2]}.
~~~\label{u-def}\eea%
Then
\bean \int d\mu^2 (\mu^2)^{-1-\eps} \to \left( {
\Delta[K,M_1,M_2]\over 4 K^2}\right)^{-\eps} \int_0^1 du~
u^{-1-\eps}.
\eean
Since ${(4\pi)^{\eps}\over (2\pi)^4\Gamma(-\eps)}\left( {
\Delta[K,M_1,M_2]\over 4 K^2}\right)^{-\eps}$ is a universal factor
on both sides (basis and amplitude) of the cut calculation, we neglect it henceforth.

Thus, we have rewritten the integral as
\bea I_n(M_1,M_2,m_1,\ldots,m_{n-2}) =
 & &\int_0^1 du~u^{-1-\eps} \int dz ~d^4\ell
~\delta^+(\ell^2)  \nonumber \\
& &(2 \ell \cdot K)
{1\over (\W \ell^2-M_1^2-\mu^2) ((\W \ell-K)^2-M_2^2-\mu^2)
\prod_{j=1}^{n-2} ((\W\ell-P_j)^2-m_{j}^2-\mu^2)},
~~~\label{n-scalar-2}\eea
where $\mu^2$ is related to $u$ through (\ref{u-def}).

%%%%%%%%%%%%%%%%%%%%%%%%%%%%%
\subsection{Kinematics and the domain of integration}
%%%%%%%%%%%%%%%%%%%%%%%%%%%%

In this subsection we determine the integration domain.

 Assuming $K^2 \neq 0$, we
choose the frame where $\vec{K}=(K,0,0,0)$, $\W \ell=(x,y,0,0)$.
Then the second cut propagator is $\W \ell -\vec{K}=(x-K,y,0,0)$.
From the on-shell conditions,
\bean  x^2-y^2= M_1^2+\mu^2,~~~~~(x-K)^2-y^2=M_2^2+\mu^2,\eean
we solve for $x$ and $y$ to find
\bea x={K^2+M_1^2-M_2^2\over 2K},~~~~~y=\pm \sqrt{
x^2-M_1^2-\mu^2}.\eea
The requirement that $y$ has a real solution is the following constraint:
\bea \mu^2\leq { \Delta[K,M_1,M_2]\over 4
K^2},~~~~\label{mu-phys}
\eea
with the definition of $\Delta[K,M_1,M_2]$ given in  (\ref{delta-cut}).
The condition (\ref{mu-phys}) also requires the right hand side to be positive,
which can be arranged by working in a region with sufficiently large cut momentum, $K > M_1 + M_2$.
Then the physical constraint (\ref{mu-phys}) restricts $u$ to lie in a unit interval:
\bea
u \in [0,1].
\eea
After using (\ref{changing}) to decompose the vector $\W \ell=\ell+ z K$,
the lightlike condition $\ell^2=0$ becomes
 $(x-z K)^2= y^2$, or
\bean  z^2 K^2-z (K^2+M_1^2-M_2^2) +(M_1^2+\mu^2) =0.\eean
Solving this equation, we find
\bea z = { (K^2+M_1^2-M_2^2)\pm \sqrt{\Delta[K, M_1, M_2]- 4 K^2
\mu^2}\over 2 K^2}.
~~~\label{twoz}
\eea
In the kinematic region of this cut, $K>0$.  Then the positive-light-cone condition
$\delta^+( \ell^2)$ is equivalent to $x-zK >0$.  Consequently, exactly one of the
two solutions (\ref{twoz}) for $z$ is selected.  Specifically,
\bea z = { (K^2+M_1^2-M_2^2)- \sqrt{\Delta[K, M_1, M_2]- 4 K^2
\mu^2}\over 2 K^2},\eea
which we rewrite as
\bea z ={ a -b \sqrt{1-u}\over 2},~~~~\label{z-sol-u}\eea
where we have defined two useful parameters:
\bea a \equiv { K^2+M_1^2-M_2^2\over K^2},~~~~b \equiv {\sqrt{\Delta[K, M_1,
M_2]}\over K^2}.~~~\label{ab-def}\eea
In the massless limit $a=b=1$.

%%%%%%%%%%%%%%%%%%
\subsection{First steps: separating the four-dimensional integral}
%%%%%%%%%%%%%%%%%%%

Now we are ready to discuss the cut integral.
Consider the two delta functions of the cut propagators,
\bean
 \delta(\W \ell^2- M_1^2-\mu^2)\delta((\W
\ell-K)^2-M_2^2-\mu^2).
\eean
Inside the integral, we perform the following manipulations:
\bean & & \delta(\W \ell^2- M_1^2-\mu^2)\delta((\W
\ell-K)^2-M_2^2-\mu^2) \\ & & = \delta(\W \ell^2- M_1^2-\mu^2)\delta(K^2-2
\W \ell \cdot K+M_1^2-M_2^2) \\ & & =  \delta(z^2 K^2+z (2\ell\cdot
K)- M_1^2-\mu^2)\delta((1-2z)K^2-2  \ell \cdot K+M_1^2-M_2^2)
\\ & & = \delta(z(1-z) K^2+z (M_1^2-M_2^2)-
M_1^2-\mu^2)\delta((1-2z)K^2-2 \ell \cdot K+M_1^2-M_2^2)
\eean
In the last step we used the second delta function to find
 $2\ell \cdot K=(1-2z)K^2+M_1^2-M_2^2$.  We will make this substitution in
the measure (\ref{changing}).
At this point the first delta function is independent of $\ell$.
The cut part of the integral now takes the form
\bean & & \int_0^1 du~u^{-1-\eps} \int dz ((1-2z)K^2+M_1^2-M_2^2)
\delta(z(1-z) K^2+z (M_1^2-M_2^2)- M_1^2-\mu^2) \\
& & \int d^4\ell \delta^+(\ell^2)\delta((1-2z)K^2-2 \ell \cdot
K+M_1^2-M_2^2).~~~~\label{2lines} \eean
The second line is the 4-dimensional phase space integration,
which can be performed in various ways, as discussed in \cite{Anastasiou:2006jv}.
 We can integrate out $z$ using the
delta-function in the first line.  Here we need to account for the factor
$${\partial \over \partial z} (z(1-z) K^2+z (M_1^2-M_2^2)-
M_1^2-\mu^2) = (1-2z)K^2 +(M_1^2-M_2^2) ,$$
which serves to cancel the first factor of the first line of (\ref{2lines}).
Finally, we arrive at the expression
\bea   \int_0^1 du~u^{-1-\eps} \int d^4\ell
\delta^+(\ell^2)\delta((1-2z)K^2-2 \ell \cdot
K+M_1^2-M_2^2),~~~\label{gen-frame} \eea
where $z$ is related to $u$ by (\ref{z-sol-u}).

%%%%%%%%%%%%%%%%%%
\section{Recursion and reduction formulas with mass}
%%%%%%%%%%%%%%%%%%%%

In this section we compute the massive analogs of the recursion and reduction formulas for master integrals presented in \cite{Anastasiou:2006jv} and derived in detail in \cite{Anastasiou:2006gt}.
It is not hard to check that the results of this section reproduce the massless results when we set all $m_i=0$ (and hence $a=b=1$ in (\ref{ab-def})).

We refer to \cite{Anastasiou:2006gt} rather than review all details of the setup here. However, we do need to remind the reader that the results for cuts of these basis integrals were derived with spinor integration, in which the massless 4-dimensional vector $\ell$ is rewritten as
\bea
\ell = t\lambda\W\la,
\eea
and the measure transforms as
\bea
\int d^4\ell ~ \delta^{(+)}(\ell^2) ~ (\bullet ) =
\int_0^{\infty}dt~t\int_{\tilde\lambda=\bar\lambda}\vev{\lambda,
d\lambda}[\tilde\lambda,d\tilde\lambda] ( \bullet ).
\eea
Here $t$ ranges over the positive real line, and $\la,\W\la$ are homogeneous spinors, also written respectively as $|\ell\rangle, |\ell]$ in many expressions involving spinor products.  The first step in spinor integration is to integrate over the variable $t$ simply by solving the delta function of the {\em second} cut propagator.  In this section, we sketch the beginnings of certain derivations by writing the integrand before and after this $t$-integration--which, we emphasize, is not true integration.

Spinor integration proceeds by writing the integrand as a total derivative plus delta functions using ``holomorphic anomaly'' formulas, and finally identifying the contributions of delta functions as residues of a complex function.  For an exposition of this technique, we refer the reader to \cite{Britto:2005ha,Britto:2006sj}.

%%%%%%%%%%%%%%%%%%%%
\subsection{Bubble}
%%%%%%%%%%%%%%%%%%%%

The cut bubble is exactly the integral described in the previous section without any additional factors.  We just need to do the four-dimensional integral.
\bean & & \int dz~ d^4\ell ~\delta^+(\ell^2) (2 \ell \cdot
K)\delta(z(1-z) K^2+z (M_1^2-M_2^2)- M_1^2-\mu^2)\delta((1-2z)K^2-2
\ell \cdot K+M_1^2-M_2^2)\\ & = & \int dz ~d^4\ell ~\delta^+(\ell^2)
((1-2z)K^2+M_1^2-M_2^2) \times \\ & & \delta(z(1-z) K^2+z (M_1^2-M_2^2)-
M_1^2-\mu^2)\delta((1-2z)K^2-2 \ell \cdot K+M_1^2-M_2^2).
\eean
After  phase space integration we are left with
\bea { (1-2z)K^2+M_1^2-M_2^2\over K^2}= b \sqrt{1-u},
\eea
where we have put in $z$ given by (\ref{z-sol-u}).
So, the cut is
\bea C[I_2(M_1,M_2;K)]= b \int_0^1 du~u^{-1-\eps}  \sqrt{1-u},
\eea
where $b$ is given in (\ref{ab-def}).
This factor of $b$ is where this expression differs from the massless case.

{\bf Recursion relation:} Now we define a sequence of functions indexed by nonnegative integers $n$:
\bea  {\rm Bub}^{(n)} & \equiv  &  \int_0^1 du~u^{-1-\eps} u^n
\sqrt{1-u}.~~~\label{I-2m-n} \eea
Note that the physical cut is related to the zeroth function by a factor of $b$:
\bea
C[I_2(M_1,M_2; K)]=
b~ {\rm Bub}^{(0)}.~~~~\label{conv-bub}
\eea
Because the definition (\ref{I-2m-n}) is identical to the massless case,
we derive exactly the same recursion formula as in \cite{Anastasiou:2006jv,Anastasiou:2006gt}:
\bea  {\rm Bub}^{(n)} & = & F_{2\to 2}^{(n)}
 {\rm Bub}^{(0)},~~~\label{I-2m-recu-3} \eea
where the form factor is
\bea F_{2\to 2}^{(n)} & = & {(-\eps)_{3 \over 2} \over (n-\eps)_{3 \over 2}}. ~~~\label{F-2n-2n} \eea

Here  $(x)_n = \Gamma(x+n)/\Gamma(x)$ is the Pochhammer symbol.

Written in a form suitable for reading master integrands, we have the result
\bea   b \int_0^1 du~u^{-1-\eps} u^n
\sqrt{1-u} =  F_{2\to 2}^{(n)} C[I_2(M_1,M_2; K)].~~~\label{bub-phys} \eea
%

%%%%%%%%%%%%%%%%%%%%%
\subsection{Triangle}
%%%%%%%%%%%%%%%%%%%%%

The integrand to start with is
\bean {\delta(\W \ell^2- M_1^2-\mu^2)\delta((\W
\ell-K_1)^2-M_2^2-\mu^2)\over ((\W \ell+K_3)^2-m_1^2-\mu^2)}.
\eean
After $t$-integration we get
\bean  - \left( (1-2z) +{ M_1^2-M_2^2\over K_1^2}\right){1\over
\gb{\ell|K_1|\ell}\gb{\ell|Q|\ell}}
= - b \sqrt{1-u} {1\over
\gb{\ell|K_1|\ell}\gb{\ell|Q|\ell}},
\eean
where
\bea Q=\left( (1-2z) +{ M_1^2-M_2^2\over K_1^2}\right) K_3+{
K_3^2+M_1^2-m_1^2+z(2K_1\cdot K_3) \over K_1^2} K_1. \eea
The four-dimensional integral gives
\bea
-{1\over \sqrt{\Delta_3}}
\left( (1-2z) +{ M_1^2-M_2^2\over
K_1^2}\right) \ln \left( { -(2 K_1\cdot Q)
+\sqrt{\Delta_3} \over -(2 K_1\cdot Q)
-\sqrt{\Delta_3}}\right),~~~\label{I3m-cut}\eea
where
\bea \Delta_3 & = & 4(( K_1\cdot Q)^2- K_1^2
Q^2).~~~~\label{Delta-3}\eea
The ingredients for (\ref{Delta-3}) are
\bean K_1\cdot Q & = & { K_1^2+M_1^2-M_2^2\over K_1^2} (K_1\cdot
K_3) + (K_3^2+M_1^2-m_1^2) \\
Q^2 & = & { (2 K_1^2 (K_3^2+M_1^2-m_1^2) + (K_1^2+M_1^2-M_2^2)
(2K_1\cdot K_3))^2- (1-u) \Delta_{3;m=0}\Delta[K_1,M_1,M_2]\over 4
(K_1^2)^3} \eean
where
\bea \Delta_{3;m=0}\equiv 4((K_1\cdot K_3)^2- K_1^2 K_3^2),
\eea
which we recognize as the signature of a triangle with massless propagators.

We can now see that
\bea \Delta_3 & = & { (1-u) \Delta_{3;m=0}\Delta[K_1,M_1,M_2] \over
(K_1^2)^2}=b^2(1-u) \Delta_{3;m=0}.~~~\label{Delta-I3m-m}\eea
It is interesting to see that $\Delta_3$
is built from the factors
 $\Delta[K_1,M_1,M_2]$, the signature of the cut,
and $\Delta_{3;m=0}$, the signature of the triangle.

Now we define
\bea Z \equiv -{ (2K_1\cdot Q) K_1^2 \over
\sqrt{\Delta_{3;m=0}\Delta[K_1,M_1,M_2]}}.~~~~\label{I3m-Z-def}\eea
Then we have
\bea
C[I_3(M_1,M_2, m_1;K_1,K_3)]
& = & \int_0^1 du
u^{-1-\eps}\left( -b \sqrt{1-u}\right){1\over
\sqrt{\Delta_3}} \ln \left( {
 {Z+\sqrt{1-u}\over Z-\sqrt{1-u}}}\right)\nonumber \\
& = &\int_0^1 du~u^{-1-\eps}
 \left(-{1\over \sqrt{\Delta_{3;m=0}}}\right)\ln \left( {
 {Z+\sqrt{1-u}\over Z-\sqrt{1-u}}}\right).~~~\label{I3m-cut-u}\eea
This is the same expression as in the massless case, but now with a different $Z$.

{\bf Recursion/reduction relation:}
We define the integrals
\bea  {\rm Tri}^{(n)}(Z) \equiv  \int_0^1 du~u^{-1-\eps}
u^n\ln \left( {Z +\sqrt{1-u}\over Z-\sqrt{1-u}
}\right),~~~\label{I-3n}\eea
where the parameter $Z$ is defined as in (\ref{I3m-Z-def}).
The physical cut integral is
\bea
C[I_{3}(M_1,M_2,m_1;K_1,
K_3)]=-{1 \over \sqrt{\Delta_{3;m=0}}} {\rm Tri}^{(0)}(Z).
~~~~\label{conv-tri}
\eea
In this case, our cut triangle functions (\ref{I-3n}) do depend on the
cut-propagator masses $M_1$ and $M_2$ via $Z$. Apart from the generalized definition of $Z$, the formula is the same as in the massless case.  Therefore, we derive the same recursion and reduction formulas as in \cite{Anastasiou:2006gt}, namely
\bea {\rm Tri}^{(n)}(Z) & = &  F_{3\to 3}^{(n)}(Z){\rm Tri}^{(0)}(Z)
+\W F_{3\to 2}^{(n)}(Z){\rm Bub}^{(0)},~~~\label{I-3n-rec-2} \eea
where the two form factors are given by
\bea F_{3\to 3}^{(n)}(Z) & = & {-\eps\over n-\eps} (1-Z^2)^n,
~~~\label{F-3n-3n}
\\
\W F_{3\to 2}^{(n)}(Z) & = &  {(-\eps)_{3 \over 2} \over n-\eps}  \sum_{k=1}^n {{2 Z}(1-Z^2)^{n-k} \over (k-\eps)_{1 \over 2}}.
 ~~~\label{F-3n-2n} \eea
These are functions of the variable $Z$, defined for a given triangle by (\ref{I3m-Z-def}).

Equation (\ref{I-3n-rec-2}) is not
yet in the most applicable form.  We return to the language of
physical cuts by including the factor $-1/\sqrt{\Delta_{3,m=0}}$
from (\ref{conv-tri}).  The recursion/reduction formula that we need
is thus:
\bea & & \int_0^1 du~u^{-1-\eps} u^n\left[
-{1\over\sqrt{\Delta_{3;m=0}}}
\ln \left( {Z +\sqrt{1-u}\over Z-\sqrt{1-u}
}\right)\right]= \nonumber \\ & &
F_{3\to 3}^{(n)}(Z) C[I_3(M_1,M_2,m_1;K_1,
K_3)]+ F_{3\to
2}^{(n)}(K_1,K_3)C[I_2(M_1,M_2;K_1)],~~~\label{I-3n-rec-3}\eea
where
\bea F_{3\to 2}^{(n)}(K_1,K_3))= -{1\over
b \sqrt{\Delta_{3;m=0}}}\W F_{3\to 2}^{(n)}(Z).\eea
%

%%%%%%%%%%%%%%%%%%%%%%%%%%%%
\subsection{Box}
%%%%%%%%%%%%%%%%%%%%%%%%%%%%

The integrand to start with is
\bean {\delta(\W \ell^2- M_1^2-\mu^2)\delta((\W
\ell-K)^2-M_2^2-\mu^2)\over ((\W \ell-P_1)^2-m_1^2-\mu^2)((\W
\ell-P_2)^2-m_2^2-\mu^2)}.
\eean
We define another useful mass-dependent parameter:
\bea
a_i \equiv {P_i^2 +M_1^2-m_i^2 \over K^2}.
\eea
After $t$-integration we get
\bean
{b \sqrt{1-u} \over K^2}{1\over \gb{\ell|Q_1|\ell}\gb{\ell|Q_2|\ell}},
\eean
where
\bea Q_i & =& (-b\sqrt{1-u})P_i+ \left( a_i- {P_i \cdot K \over K^2}
(a-b\sqrt{1-u})\right) K.
~~~~\label{qi}
\eea
Now the procedure is the same as in the massless case, but with this
more general definition of $Q_i$.

Define
\bea
R_i & = & \left. Q_i \right|_{u=0} \\
&=& -b P_i+ \left( a_i- {P_i \cdot K \over K^2}
(a-b )\right) K.
\eea
The physical cut is
\bea & & C[I_4(M_1,M_2,m_1,m_2;K,P_1,P_2)  =~~~\label{box-mass}
\\
& &  \int_0^1 du~u^{-1-\eps} {b\over 2K^2}{1\over \sqrt{B-A u}}\ln
\left({D-C u +\sqrt{1-u}\sqrt{B-A u}\over D-C u -\sqrt{1-u}\sqrt{B-A
u}} \right),\nonumber \eea
where
\bea
 & A = -{b^4\over K^2}\det\left( \begin{array}{ccc} P_1^2~~ & P_1\cdot
P_2~~ & P_1\cdot K \\ P_1\cdot P_2~~ & P_2^2~~ & P_2\cdot K \\
P_1\cdot K~~ & P_2\cdot K~~ &
K^2\end{array}\right),~~~~~~~~ & C= {b^2\over K^2}\det\left(
\begin{array}{cl}  P_1\cdot P_2~~ & P_1\cdot K \\ P_2\cdot K~~ & K^2
\end{array}\right), \nonumber \\
& B = -\det \left( \begin{array}{cc}
R_1^2 & R_1\cdot R_2 \\
R_1\cdot R_2 &
R_2^2
\end{array}\right),~~~~~~~~~~~
& D = R_1\cdot R_2.
~~~~\label{Para-3}
\eea
Here again, the form of (\ref{box-mass}) differs from the one in massless
case only by the factor of $b$.

{\bf Recursion/reduction relation:} We define
\bea   {\rm Box}^{(n)}(A,B,C,D)  \equiv   \int_0^1
du~u^{-1-\eps}\nonumber  {u^n\over \sqrt{B-A u}} \ln
\left( {D-C u+ \sqrt{1-u}\sqrt{B-A u}\over D-C u-
\sqrt{1-u}\sqrt{B-A u}}\right). ~~~\label{4m-n} \eea
The physical cut is related to the zeroth function in (\ref{4m-n}) by
\bea C[I_{4}(M_1,M_2,m_1,m_2;K,P_1,P_2)] =  {b\over
2K^2} {\rm Box}^{(0)}(A,B,C,D) \eea
if $A,B,C,D$ are defined as in (\ref{Para-3}).

 In Table
(\ref{Box-cut-table}) we have listed which kinds of triangles a box
with given cut would reduce to (the propagator mass must also be
correctly identified):
\bea
\begin{array}{|c|c|c|c|c|} \hline
Box~Cut~K & ~~~P_1~~~ & ~~~P_2~~~ & Triangle~One's~(K_1, K_3) &
Triangle~Two's~(K_1, K_3)
\\ \hline K_1 & K_{12} & -K_4 &  (K_1, K_{34}) & (K_1, K_4) \\
\hline K_2 & K_{23} & -K_1 & (K_2, K_{41}) & (K_2, K_1) \\
\hline K_3 & K_{34} & - K_2 & (K_3, K_{12}) & (K_3, K_2) \\
\hline K_4 & K_{41} & - K_3 & (K_4, K_{23}) & (K_4, K_3) \\
\hline K_{12} & K_1  & - K_4 & (K_{34}, K_2) & (K_{12}, K_4) \\
\hline K_{23} & K_2 & - K_1 & (K_{41}, K_3) & (K_{23}, K_1) \\
\hline
\end{array}~~~\label{Box-cut-table}
\eea
Boxes are related to triangles and bubbles in the reduction formulas.
In these relations we make use of a quantity combining variables of the box and the associated triangles to Table (\ref{Box-cut-table}):
\bea
C_{Z_i} \equiv (Z_i^2-1)C +D.
 \eea
\bea & & {\rm Box}^{(n)}(A,B,C,D) =  F_{4\to 4}^{(n)}(A,B){\rm
Box}^{(0)}(A,B,C,D)+ \W F_{4\to
3}^{(n)}(A,B,C,D;Z_1){\rm Tri}^{(0)}(Z_1)\nonumber\\
& & ~~~~+ \W F_{4\to 3}^{(n)}(A,B,C,D;Z_2) {\rm Tri}^{(0)}(Z_2)+ \W
F_{4\to 2}^{(n)}(A,B,C,D;Z_i) {\rm Bub}^{(0)},~~~\label{I-4n-rec-2}
\eea
where the form factors are given by
\bea F_{4\to 4}^{(n)}(A,B) & = &
 { (-\eps)_{1 \over 2} \over (n-\eps)_{1 \over 2}} \left({B \over A}\right)^n,
~~~\label{F-4n-4n}
\\
\W F_{4\to 3}^{(n)}(A,B,C,D;Z_i) & = &
-{C_{Z_i}\over  (n-\eps)_{1 \over 2}A ~Z_i} \sum_{k=1}^n
{(k-1-\eps)_{1 \over 2}}\left({B \over
A}\right)^{n-k} F^{(k-1)}_{3\to 3}(Z_i),~~~\label{F-4n-3n}
\\
\W F_{4\to 2}^{(n)}(A,B,C,D;Z_i)  & = &
-{\Gamma(n-\eps) \over \Gamma(n+1/2-\eps)}
{1\over A} \nonumber \\
& & \hspace{-1.2cm} \times
 \sum_{k=1}^n { (k-1-\eps)_{1 \over 2}}
\left({B \over A}\right)^{n-k}
\left( {C_{Z_1}\over Z_1}\W F^{(k-1)}_{3\to 2}(Z_1) + {C_{Z_2}\over
Z_2}\W F^{(k-1)}_{3\to 2}(Z_2) \right). ~~~\label{F-4n-2n} \eea

Again (\ref{I-4n-rec-2}) is not the final formula we are after. To
get the proper physical result for identifying integrands, we need to replace the kinematic
factor ${b/ 2K^2}$. The result is
\bea & & \int_0^1 du~ u^{-1-\eps} u^n \left[{b\over 2K^2 \sqrt{ B -
A u}} \ln \left( {D - C u+ \sqrt{1-u}\sqrt{ B - A u}\over D - C u-
\sqrt{1-u}\sqrt{ B - A u}}\right)\right]  =  \nonumber \\ & & ~~~~~~~~~
F_{4\to 4}^{(n)}(A,B)
C[I_4(M_1,M_2,m_1,m_2;K,P_1, P_2)] \nonumber \\ & & ~~~~~~~~~
+  \sum_{i=1}^2 F_{4\to
3}^{(n)}(A,B,C,D;Z_i) C[I_{3}(M_1,M_2,m_1^{(i)};K_1^{(i)},K_3^{(i)})] \nonumber \\ & & ~~~~~~~~~
+ F_{4\to
2}^{(n)}(A,B,C,D;Z_i) C[I_2(M_1,M_2;K)], ~~~\label{I-4n-rec-3} \eea
where for the triangles, $K_1^{(i)}$ and $K_3^{(i)}$ are given by Table
(\ref{Box-cut-table}), the mass $m_1^{(1)}$ must also be interpreted correctly according to the same table, and the form factors are
\bea F_{4\to 3}^{(n)}(A,B,C,D;Z_i)& = & -{b\sqrt{\Delta_{3}^{(i)}}\over
  2K^2} ~ \W
F_{4\to 3}^{(n)}(A,B,C,D;Z_i), \nonumber \\ F_{4\to
2}^{(n)}(A,B,C,D;Z_i)& = &  {1\over
2K^2} ~ \W F_{4\to 2}^{(n)}(A,B,C,D;Z_i).
\eea
%

%%%%%%%%%%%%%%%%%%%
\subsection{Pentagon}
%%%%%%%%%%%%%%%%%%%

The integrand is
\bean {\delta(\W \ell^2- M_1^2-\mu^2)\delta((\W
\ell-K)^2-M_2^2-\mu^2)\over ((\W \ell-P_1)^2-m_1^2-\mu^2) ((\W
\ell-P_2)^2-m_2^2-\mu^2) ((\W \ell-P_3)^2-m_3^2-\mu^2)}.\eean
The physical cut is
\bean
C[I_5(M_1,M_2,m_1,m_2,m_3;K,P_1,P_2,P_3)]=
\int_0^1 du~u^{-1-\eps}
 \int \vev{\ell~d\ell}[\ell~d\ell] { b \sqrt{1-u}
\gb{\ell|K|\ell} \over (K^2)^2
\gb{\ell|Q_1|\ell}\gb{\ell|Q_2|\ell}\gb{\ell|Q_3|\ell}}, \eean
where
\bea
 Q_i= -\left( b \sqrt{1-u}
\right) P_i+{ P_i^2+M_1^2-m_i^2-2z(K\cdot P_i) \over K^2} K.
~~~~\label{otherQu} \eea

The total integral (apart from the universal prefactor) is
\bea b \int_0^1 du~u^{-1-\eps}
 \sqrt{1-u} \int \vev{\ell~d\ell}[\ell~d\ell]
{ \gb{\ell|K|\ell} \over (K^2)^2
\gb{\ell|Q_1|\ell}\gb{\ell|Q_2|\ell}\gb{\ell|Q_3|\ell}}.
 \eea

Apart from the factor of $b$ and the modified definition of $Q_i$,
the analysis proceeds as in  the massless case \cite{Anastasiou:2006gt}. Thus we can cite the
result directly as
\bea & & C[I_5(M_1,M_2,m_1,m_2,m_3;K,P_1,P_2,P_3)] = -b\int_0^1 du
u^{-1-\eps}
 {\sqrt{1-u} \over (K^2)^2}~~~\label{Pentagon-gen} \\ &  &
\left({ S[Q_3, Q_2, Q_1,K]\over 4\sqrt{(Q_3\cdot Q_2)^2 -Q_3^2
Q_2^2}} \ln{  Q_3 \cdot Q_2 - \sqrt{  (Q_3\cdot Q_2)^2 -Q_3^2
Q_2^2}\over  Q_3 \cdot Q_2 + \sqrt{  (Q_3\cdot Q_2)^2 -Q_3^2
Q_2^2}}\right. \nonumber \\ & & + { S[Q_3, Q_1, Q_2,K]\over
4\sqrt{(Q_3\cdot Q_1)^2 -Q_3^2 Q_1^2}} \ln{ Q_3 \cdot Q_1 - \sqrt{
(Q_3\cdot Q_1)^2 -Q_3^2 Q_1^2}\over  Q_3 \cdot Q_1 + \sqrt{
(Q_3\cdot Q_1)^2 -Q_3^2 Q_1^2}}  \nonumber \\ & & \left.+ { S[Q_2,
Q_1, Q_3,K]\over 4\sqrt{(Q_2\cdot Q_1)^2 -Q_2^2 Q_1^2}} \ln{  Q_2
\cdot Q_1 - \sqrt{  (Q_2\cdot Q_1)^2 -Q_2^2 Q_1^2}\over  Q_2 \cdot
Q_1 + \sqrt{ (Q_2\cdot Q_1)^2 -Q_2^2 Q_1^2}}\right),\nonumber \eea
where $S[Q_3, Q_2, Q_1,K]$ is a rational function defined as follows:
\bea  & & S[Q_2, Q_1, Q_3,K]  =  {T_1\over T_2},~~~\label{Func-S} \eea
with
\bea
T_1 = -8 \det \left( \begin{array}{lcr} Q_3 \cdot K & Q_2 \cdot K & Q_1 \cdot K\\
Q_2 \cdot Q_3 & Q_2^2 & Q_2 \cdot Q_1 \\ Q_1 \cdot Q_3 & Q_2 \cdot Q_1 &
Q_1^2
\end{array} \right), ~~~~
T_2 = -4 \det \left( \begin{array}{lcr} Q_3^2 & Q_2 \cdot Q_3 & Q_1 \cdot Q_3\\
Q_2 \cdot Q_3 & Q_2^2 & Q_2 \cdot Q_1 \\ Q_1 \cdot Q_3 & Q_2 \cdot Q_1 &
Q_1^2
\end{array} \right).~~~~\label{T1-T2}
\eea

Let us make a few comments on the behavior of pentagon cuts.
 There are three
terms.
Each term looks like a box signature multiplied by the factor
 $S[\bullet]/( 2 K^2)$.
It is significant that each function $S[\bullet]$ has the same denominator $T_2[\bullet]$, which does not depend on the order of the first three arguments.  This can be considered as another  signature
of a cut-pentagon integral.
This feature makes the reduction simple.
Where we see a factor of $u^n$,
 we just
need to write $u^n S= P(u)+  A \sum_{i=1}^3 S_i[\bullet]$ where
$P(u)$ is a polynomial in $u$ and $A$ is constant in $u$.\footnote{It
is essential that because all three $S_i[\bullet]$ have same
denominator $T_2$, after reduction we have the same $A$ multiplying all
three of the $S_i[\bullet]$.} The $A$ term will be the pentagon
coefficient, while $P(u)$ indicates reduction to boxes. In the
massless case, pentagons contribute to terms of ${\cal O}(\epsilon)$,
so they can be neglected. However, in cases where
propagators are massive or we wish to compute to higher orders in $\eps$, their contribution must be
included.

%%%%%%%%%%%%%%%%%%%%%%%%%%
\section{Formulas for coefficients from double cuts}
%%%%%%%%%%%%%%%%%%%%%%%%%%%

Systematic extraction of coefficients of master integrals from four-dimensional spinor integration techniques has been described in \cite{Britto:2005ha,Britto:2006sj}, building on earlier techniques reviewed in \cite{Cachazo:2005ga,Dixon:2005cf}.   When applied to a specific amplitude in practice, there are choices to make regarding how to ``split'' the integrand in partial fractions as well as choosing arbitrary spinors on which the final answer does not depend.

In this section we present canonical choices to further simplify the method, aiming to automatize the heart of this procedure to allow for easy implementation into a computer program.

For brevity, we do not review the entire spinor integration technique here; instead we refer the reader to the explanations given in \cite{Britto:2005ha,Britto:2006sj}.
The steps that concern us here come after the
coordinate change and $t$-integration mentioned at the beginning of Section 3.  At this point we have an integrand whose terms are rational functions of spinor products and homogeneous in the spinor integration variables.
The following steps are to split the integrand into partial fractions (using Schouten identities), followed by identification of master integrals and integration over a single Feynman parameter.  Finally we apply the holomorphic anomaly to complete the spinor integration by extracting residues.

Here we are able to give explicit algebraic functions for
coefficients. In the appendix we  give a more detailed discussion of
how to evaluate these functions in practice.

A summary of the results  this section may be found in Section 4.4.

%%%%%%%%%%%%%%%%
\subsection{Canonical splitting}
%%%%%%%%%%%%%%%%%%

Recall that our starting point is the product of the two (on-shell) tree-level amplitudes from each side of the unitarity cut. 
In spinor notation, this integrand takes the general form\footnote{In this
section we deal specifically with the case of massless propgators.  The generalization to the 
massive case is straightforward, as in the previous sections.}
\bean \sum { C\prod_{i=1}^{n_1} \gb{\W a_i|\ell|\W b_j} \over
\prod_{j=1}^{n_2} ((\W\ell-P_j)^2-\mu^2)}.\eean
Here $C$ is some expression that does not depend on any integration variable, and
we have used the relation (\ref{changing}), $\W \ell=\ell+z K$, to rewrite the  numerator. 
As we emphasized in the beginning of Section 3, the $t$-integration is trivial because of the second delta function,
$\delta((1-2z)K^2-2 \ell \cdot K)$.
Thus we can write the result of this step immediately: 
\bean { (1-2z) K^2\over \gb{\ell|K|\ell}^2}\sum \left(-{(1-2z)K^2\over \gb{\ell|K|\ell}}\right)^{n_1}{ 
\gb{\ell|K|\ell}^{n_2} C \prod_{i=1}^{n_1} \vev{\W a_i~\ell}[\ell~\W
b_i] \over \prod_{j=1}^{n_2} \gb{\ell|Q_j|\ell}}.\eean
So we see that, after integrating over $t$, the
integrand for phase space integration is a sum of terms of the
following form:
\bea I_{term}={ G(\la) \prod_{j=1}^{n+k-2} [a_j~\ell]\over
\gb{\ell|K|\ell}^n
\prod_{i=1}^{k}\gb{\ell|Q_i|\ell}}.~~\label{gen-form}\eea
Here $G(\la)$ is some monomial in the holomorphic spinor only; hence it factorizes as $\prod \vev{\ell~c_i}$ times a constant in $\ell$.  The functions $a_j$ may depend on $\la$ as well, for example as $[a_j|=\langle \la|Q|$. But $a_j$ is the quantity we will need in order to split the term further and extract the residues from multiple poles.

We aim to describe all coefficients in terms of the quantities $K,Q_i, a_j$ and $G(\la)$.

The term (\ref{gen-form}) depends as well on $n$ and $k$.  The exponent $n$ is related to the type of master integral involved.  Terms with $n=0$ contribute only to  boxes and pentagons.  Terms with $n=1$ contribute to triangles in addition, and terms with $n \geq 2$ contribute to all of these plus bubbles.  If $n=0$ we can multiply both numerator and denominator by $\gb{\ell|K|\ell}$, so we shall  always assume $n\geq 1$.

The first step in our program is to isolate poles by splitting $I_{term}$ using the following partial fraction spinor identity:
\bea {[a~\ell]\over \gb{\ell|Q_1|\ell}\gb{\ell|Q_2|\ell}}
={\tgb{a|Q_1|\ell}\over \vev{\ell|Q_2
Q_1|\ell}\gb{\ell|Q_1|\ell}}+{\tgb{a|Q_2|\ell}\over \vev{\ell|Q_1
Q_2|\ell}\gb{\ell|Q_2|\ell}}.~~\label{Ge-split}\eea
Because there are different factors in the denominator, there are different sequences of splitting leading to different but equivalent expressions.
Here we choose a canonical sequence:

(1) First we keep $\gb{\ell|K|\ell}^n$ untouched and split
the factors $\gb{\ell|Q_i|\ell}$ among themselves. At the end of this step each term takes the form ${\W G(\la)\over \gb{\ell|K|\ell}^n\gb{\ell|Q_i|\ell}}$.

(2) Then we split $\gb{\ell|K|\ell}$ from $\gb{\ell|Q_i|\ell}$ as often as necessary
to get two types of terms, ${\W G_m(\la)\over \gb{\ell|K|\ell}^m}$
and ${\W F_i(\la)\over \gb{\ell|K|\ell}\gb{\ell|Q_i|\ell}}$.

(3) Finally we will be left with
\bea  { G(\la) \prod_{j=1}^{n+k-2} [a_j~\ell]\over
\gb{\ell|K|\ell}^n \prod_{i=1}^{k}\gb{\ell|Q_i|\ell}} & = &
\sum_{k=2}^n G_k(\la) { \prod_{j=1}^{k-2} [b_j~\ell]\over
\gb{\ell|K|\ell}^k} +\sum_{i=1}^k F_i(\la) {1\over
\gb{\ell|K|\ell}\gb{\ell|Q_i|\ell}}.~~\label{Expa} \eea
Our task is to find expressions for $G_k(\la)$,$F_i(\la)$ and $b_j$ in
the most compact and simple form.  The result is
\bea F_i(\la) & = &  \left( {G(\la)\prod_{s=1}^{n+k-2}
\tgb{a_s|Q_i|\ell}\over \vev{\ell|K Q_i|\ell}^{n-1}\prod_{t=1,t\neq
i}^{k} \vev{\ell|Q_t Q_i|\ell}}\right),~~~~\label{F-i-coe}\\
G_p(\la) & = & \sum_{i=1}^k  {G(\la)\prod_{s=1}^{k-1}
\tgb{a_s|Q_i|\ell}\over \prod_{t=1,t\neq i}^{k} \vev{\ell|Q_t
Q_i|\ell}}  { \prod_{l=k}^{n-k-p} \tgb{a_l|K|\ell}\over
\vev{\ell|Q_i K|\ell}^{n+1-p}},~~p=2,...,n~~~~\label{G-p-coe}\\
b_j & = & a_{j+n+k-p}.~~~\label{b-j-assign}\eea
 While $F_i(\la)$ finds its  simplest and most compact form in
(\ref{F-i-coe}),  $G_p(\la)$ in (\ref{G-p-coe}) may  not be the
simplest. For example, it can be shown that all terms with $1/
\gb{\ell|K|\ell}^n$ can be summed into a single term.
In general there are several terms with
 $1/\gb{\ell|K|\ell}^a$ with $2\leq a<n$.
As we discuss in Section 5.3, if we know that $n\leq 4$, we can work them out explicitly.
However, in Section 4.3 we shall use a slightly different method to deal
with these rational terms.

The identities necessary for carrying out steps (1) and (2) are the following,
which may be proved by induction, making use of the basic splitting identity (\ref{Ge-split}).
\bea  {\prod_{j=1}^{k-1}[a_j~\ell]\over \prod_{i=1}^k
\gb{\ell|Q_i|\ell}}=\sum_{i=1}^k {1\over
\gb{\ell|Q_i|\ell}}{\prod_{j=1}^{k-1} \tgb{a_j|Q_i|\ell}\over
\prod_{m=1,m\neq i}^k \vev{\ell|Q_m
Q_i|\ell}}~~\label{Split-ob-1}\eea
\bea {\prod_{j=1}^{n-1} [a_j~\ell]\over \gb{\ell|K|\ell}^{n}
\gb{\ell|Q|\ell}}= {\prod_{j=1}^{n-1}\tgb{a_j|Q|\ell}\over
\vev{\ell|KQ|\ell}^{n-1}} {1\over \gb{\ell|K|\ell} \gb{\ell|Q|\ell}}
+\sum_{p=0}^{n-2} (-)^{n-p}
{
\prod_{j=1}^{n-p-2}\tgb{a_j|Q|\ell}
\tgb{a_{n-p-1}|K|\ell}
\prod_{t=n-p}^{n-1}[a_t~\ell]
\over
\gb{\ell|K|\ell}^{p+2}
\vev{\ell|Q K|\ell}^{n-p-1}}~~~\label{Split-ob-2}\eea
One necessary condition for the form (\ref{Split-ob-1}) is that  all the $Q_i$ are different, which is satisfied for generic momenta.

%%%%%%%%%%%%%%%%%%%%%%%%%%%%%%%%%%
\subsection{Box and triangle coefficients}
%%%%%%%%%%%%%%%%%%%%%%%%%%%%%%%%%%

In this subsection we derive a closed expression for box coefficients and systematize the derivation of triangle coefficients.

As we know from Section 3.3, box integrals and their coefficients may be labeled by momenta $P_i,P_j$ along with the cut momentum $K$.  From these momenta we have defined $Q_i,Q_j$ as in (\ref{qi}).

The terms we need to consider are of the following form:
\bean I^{(i)} &=& \int F_i(\la) {1\over \gb{\ell|K|\ell} \gb{\ell|Q_i|\ell}}\\ & = &
\int_0^1 dx \int \vev{\ell~d\ell}[d\ell~\partial_\ell]\left({
F_i(\la) [\eta~\ell] \over \gb{\ell|R|\ell}
\gb{\ell|R|\eta}}\right),~~~~R=x Q_i+(1-x)K. \eean
As discussed in the previous subsection, there will be two types of poles:  those from  $\vev{\ell|K Q|\ell}$  contribute to triangles, and those from $\vev{\ell|Q_j Q_i|\ell}$ contribute to boxes.  Let us see precisely how these arise.

We can construct two massless momenta from $Q_i$ and $K$:
\bea
P_{1,2}^{(i)}=Q_i+ x_{1,2}^{(i)} K,~~\label{defnullqk}
\eea
where
\bea x_{1,2}^{(i)}& = &{-2 Q_i\cdot K\pm \sqrt{\Delta^{(i)}}\over 2
K^2},~~~~~~\Delta^{(i)}=(2 Q_i\cdot K)^2-4 Q_i^2 K^2.
\eea
We make the choice $\eta=P_1^{(i)}$ and  find that
\bea
I^{(i)} =
\int_0^1 dx \int \vev{\ell~d\ell}[d\ell~\partial_\ell]\left({
F_i(\la) [P_1^{(i)}~\ell] \over
\vev{\ell~P_2^{(i)}}[P_2^{(i)}~P_1^{(i)}]}{ (x_1^{(i)}-x_2^{(i)})\over \gb{\ell|R|\ell}
\left( x { (x_1^{(i)}+1) -1 }  \right)}\right).~~~\label{Int-exp-1}\eea
The poles in this expression come from $F_i(\la)$ and $\vev{\ell~P_2^{(i)}}$.  Note as well that since we have chosen $\eta=P_1^{(i)}$, the factor $[P_1^{(i)}~\ell]$ appears in the numerator and therefore there is {\em no contribution} from the pole at $\vev{\ell~P_1^{(i)}}$, which one might naively expect from a factor of  $\vev{\ell|K Q_i|\ell}$ in the  denominator of $F_i(\la)$.

Apply a partial fraction expansion to the $x$-dependent factors in
the denominator of the integral.  The result is
\bea
I^{(i)} =
\int_0^1 dx \int
\vev{\ell~d\ell}[d\ell~\partial_\ell]\left(
{F_i(\la) \over
\vev{\ell| Q_i K|\ell}}\left(
-{  (x_1^{(i)}+1)\over x  (x_1^{(i)}+1)-1}
+{\gb{\ell|Q_i-K|\ell}\over
x\gb{\ell|Q_i-K|\ell}+\gb{\ell|K|\ell}}
\right)\right),~~~\label{Int-exp-2} \eea
where we have used the fact that

\bea \vev{\ell|Q_i K|\ell}=
{\vev{\ell~P_1^{(i)}}[P_1^{(i)}~P_2^{(i)}]\vev{\ell~P_2^{(i)}}\over
(x_1^{(i)}-x_2^{(i)})}.~~~\label{vev-exp}\eea

{\bf First term of (\ref{Int-exp-2}):} Let us consider the two
terms of (\ref{Int-exp-2}) separately. The first is equal to
\bean I_1^{(i)}& \equiv &\int_0^1 dx \int
\vev{\ell~d\ell}[d\ell~\partial_\ell]
\left(
{F_i(\la) \over
\vev{\ell| Q_i K|\ell}}\left(
-{  (x_1^{(i)}+1)\over x  (x_1^{(i)}+1)-1}
\right)\right)
 \\
& = & \int \vev{\ell~d\ell}[d\ell~\partial_\ell]\left({F_i(\la)
\over \vev{\ell| Q_i K|\ell}}( -\ln (-x_1^{(i)}))\right)\\
& = & \int
\vev{\ell~d\ell}[d\ell~\partial_\ell]\left({G(\la)\prod_{s=1}^{n+k-2}
\tgb{a_s|Q_i|\ell}\over \vev{\ell|K Q_i|\ell}^{n}\prod_{t=1,t\neq
i}^{k} \vev{\ell|Q_t Q_i|\ell}}\ln (-x_1^{(i)})\right).\eean
In the last line we made the substitution (\ref{F-i-coe}).

To complete the calculation for $I_1^{(i)}$, we need to take residue
of all poles except the one $\vev{\ell~P_1^{(i)}}$ from factor
$\vev{\ell|K Q_i|\ell}$. This seems difficult to do directly. However, notice
that  the whole expression is holomorphic. Using the result that the sum
of residues of all poles of a holomorphic function is zero, we get
\bea I_1^{(i)} = -\left.\left({G(\la)\prod_{s=1}^{n+k-2}
\tgb{a_s|Q_i|\ell}\over \vev{\ell|K Q_i|\ell}^{n}\prod_{t=1,t\neq
i}^{k} \vev{\ell|Q_t Q_i|\ell}}\ln
(-x_1^{(i)})\right)\right|_{residue~of~\vev{\ell~P_1^{(i)}}}.~~~\label{IKQi-2}\eea
These residues contribute to triangle coefficients. In the appendix
we show how to evaluate the residue of a multiple pole. It can be seen that the
expression (\ref{IKQi-2}) defines the algebraic function for the
coefficients with input $a_i$ and $Q_i$.

{\bf Second term of (\ref{Int-exp-2}):} The second term of
(\ref{Int-exp-2}) is defined by
\bea I_2^{(i)}& \equiv &\int_0^1 dx \int
\vev{\ell~d\ell}[d\ell~\partial_\ell]\left({F_i(\la) \over
\vev{\ell| Q_i K|\ell}} {\gb{\ell|Q_i-K|\ell}\over
x\gb{\ell|Q_i-K|\ell}+\gb{\ell|K|\ell}}\right) \\
&= &\int_0^1 dx \int \vev{\ell~d\ell}[d\ell~\partial_\ell]\left(
-{G(\la)\prod_{s=1}^{n+k-2} \tgb{a_s|Q_i|\ell}\over \vev{\ell|K
Q_i|\ell}^{n}\prod_{t=1,t\neq i}^{k} \vev{\ell|Q_t
Q_i|\ell}}{\gb{\ell|Q_i-K|\ell}\over
x\gb{\ell|Q_i-K|\ell}+\gb{\ell|K|\ell}}\right).~~~~\label{Int-exp-22}
\eea
Again, we do not take the residue of $\vev{\ell~P_1^{(i)}}$ from
the factor $\vev{\ell|K Q_i|\ell}$.

There are two kinds of poles in (\ref{Int-exp-22}): one is
a simple pole from $\vev{\ell|Q_j
Q_i|\ell}$ and one is a possible multiple pole from $\vev{\ell~P_2^{(i)}}^n$ within $\vev{\ell|K
Q_i|\ell}^{n}$.  The former contributes to boxes; the latter contributes to triangles.  Let us consider the multiple poles first.

When we replace  $|\ell]\to
|P_2^{(i)}]$, the spinor dependence cancels out in the integrand, giving
\bean {\gb{\ell|Q_i-K|\ell}\over
x\gb{\ell|Q_i-K|\ell}+\gb{\ell|K|\ell}} & \to &
 { { (x_2^{(i)}+1)}\over x {
(x_2^{(i)}+1) }-1}.
\eean
We integrate over the Feynman parameter and find that the residue is
\bea \left. I_2^{(i)} \right|_{\vev{\ell~P_2^{(i)}}} =
\left.\left(-{G(\la)\prod_{s=1}^{n+k-2} \tgb{a_s|Q_i|\ell}\over
\vev{\ell|K Q_i|\ell}^{n}\prod_{t=1,t\neq i}^{k} \vev{\ell|Q_t
Q_i|\ell}}\ln
(-x_2^{(i)})\right)\right|_{residue~of~\vev{\ell~P_2^{(i)}}}.~~~\label{IKQi-1-multi}\eea
Like (\ref{IKQi-2}), it will contribute to triangles.

Now we move to the simple poles in (\ref{Int-exp-22}). For these
simple poles we can evaluate the integral over the Feynman parameter
first and get
\bea \left. I_2^{(i)}\right|_{simple~poles} & = & \int
\vev{\ell~d\ell}[d\ell~\partial_\ell]\left(
{G(\la)\prod_{s=1}^{n+k-2} \tgb{a_s|Q_i|\ell}\over \vev{\ell|K
Q_i|\ell}^{n}\prod_{t=1,t\neq i}^{k} \vev{\ell|Q_t Q_i|\ell}}\ln
{\gb{\ell|K|\ell}\over \gb{\ell|Q_i|\ell}}\right).~~~\label{i1i-simple}
\eea
Now we need to  compute the residue from the poles in $\vev{\ell|Q_j
Q_i|\ell}$.  Assume without loss of generality that $i<j$, and
construct two massless momenta as
\bea P^{(ij)}_{1,2} & = & Q_j+ y^{(ij)}_{1,2} Q_i,~~~~~~~(i<j)
\eea
where
\bea y_{1,2}^{(ij)}& = &{-2 Q_i\cdot Q_j\pm
\sqrt{\Delta^{(ij)}}\over 2
Q_i^2},~~~\Delta^{(ij)}=(2 Q_i\cdot Q_j)^2-4 Q_i^2 Q_j^2.
~~~\label{nullforqq}
\eea
To simplify our expressions, we define the following function:
\bea
F_{i,j}(\ell) = {G(\la)\prod_{s=1}^{n+k-2} \tgb{a_s|Q_i|\ell}\over
\vev{\ell|K Q_i|\ell}^{n}\prod_{t=1,t\neq i,j}^{k} \vev{\ell|Q_t
Q_i|\ell}}.~~~~\label{shorthandFij}
\eea
To sum the contributions from these two simple poles
$\vev{\ell~P_1^{(ij)}}$ and $\vev{\ell~P_2^{(ij)}}$,
notice that the $P_{1,2}^{(ij)}$ differ only by a sign in front of
the square root. Thus we can expand
\bea F_{i,j}(P_1^{(ij)})=
F_{i,j}^{(S)}+F_{i,j}^{(A)},~~~F_{i,j}(P_2^{(ij)})=
F_{i,j}^{(S)}-F_{i,j}^{(A)}.
\eea
Putting this back into the expression (\ref{i1i-simple}) for the residue, it is
straightforward to derive that the contribution from these two poles
in $I_2^{(i)}$ is
\bean
 -{1\over
\sqrt{\Delta^{(ij)}}}\left( F_{i,j}^{(S)}\ln
{\gb{P_1^{(ij)}|K|P_1^{(ij)}}\over \gb{P_1^{(ij)}|Q_i|P_1^{(ij)}}}
{\gb{P_2^{(ij)}|Q_i|P_2^{(ij)}}\over
\gb{P_2^{(ij)}|K|P_2^{(ij)}}}+F_{i,j}^{(A)}\ln
{\gb{P_1^{(ij)}|K|P_1^{(ij)}}\over \gb{P_1^{(ij)}|Q_i|P_1^{(ij)}}}
{\gb{P_2^{(ij)}|K|P_2^{(ij)}}\over
\gb{P_2^{(ij)}|Q_i|P_2^{(ij)}}}\right). \eean

To proceed further, we
notice that the same simple-pole factor  $\vev{\ell|Q_j Q_i|\ell}$ shows up in
the $I^{(j)}$ term (or $F_j(\la)$ in (\ref{Expa})). We can do
a similar calculation to get residues in $I_2^{(j)}$ in terms the
function $F_{j,i}(\la)$.
One can easily check that
$F_{j,i}(P_{1,2}^{(ij)})=F_{i,j}(P_{1,2}^{(ij)}),$
by noticing that
\bea F_{j,i}(P_{1,2}^{(ij)})= {G(P_{1,2}^{(ij)})\prod_{s=1}^{n+k-2}
[a_s~P_{2,1}^{(ij)}]\over
\gb{P_{1,2}^{(ij)}|K|P_{2,1}^{(ij)}}^{n}\prod_{t=1,t\neq j,i}^{k}
\gb{P_{1,2}^{(ij)}|Q_t|P_{2,1}^{(ij)}}}=F_{i,j}(P_{1,2}^{(ij)}).~~~\label{box-result}
\eea
Therefore the sum of contributions from residues associated to
$\vev{\ell|Q_j Q_i|\ell}=-\vev{\ell|Q_i Q_j|\ell}$ in $I_2^{(i)}$
and $I_2^{(j)}$ is
\bea  & & -{1\over \sqrt{\Delta^{(ij)}}}\left( F_{i,j}^{(S)}\ln
{\gb{P_1^{(ij)}|Q_j|P_1^{(ij)}}\over \gb{P_1^{(ij)}|Q_i|P_1^{(ij)}}}
{\gb{P_2^{(ij)}|Q_i|P_2^{(ij)}}\over
\gb{P_2^{(ij)}|Q_j|P_2^{(ij)}}}+F_{i,j}^{(A)}\ln
{\gb{P_1^{(ij)}|Q_j|P_1^{(ij)}}\over \gb{P_1^{(ij)}|Q_i|P_1^{(ij)}}}
{\gb{P_2^{(ij)}|Q_j|P_2^{(ij)}}\over
\gb{P_2^{(ij)}|Q_i|P_2^{(ij)}}}\right)\\ & = & -{1\over
\sqrt{\Delta^{(ij)}}}\left( F_{i,j}^{(S)}\ln { y_1^{(ij)}\over
y_2^{(ij)}}+F_{i,j}^{(A)}\ln { Q_j^2\over Q_i^2} \right)
\\ & = & -{1\over
\sqrt{\Delta^{(ij)}}}\left( {F_{i,j}(P_{1}^{(ij)})+
F_{i,j}(P_{2}^{(ij)})\over 2}\ln{-2 Q_i\cdot Q_j+
\sqrt{\Delta^{(ij)}}\over -2 Q_i\cdot Q_j- \sqrt{\Delta^{(ij)}}}+
{F_{i,j}(P_{1}^{(ij)})- F_{i,j}(P_{2}^{(ij)})\over 2}\ln {
Q_j^2/K^2\over Q_i^2/K^2} \right).~~~~\label{qqpoles}\eea
This is the result we are looking for.  The first term is the box
contribution. To fix sign conventions, recall that the double cut is
given by\footnote{In this part, we compare to the case where
propagators are massless. For the massive case, we
need to adjust the definition of $Q_i$ and include a factors of $b$, as explained in section 3.}
\bean & & {(1-2z)\over K^2}\int \vev{\ell~d\ell}[\ell~d\ell] {
1\over \gb{\ell|Q_1|\ell}\gb{\ell|Q_2|\ell}}
\\ & &
= {(1-2z)\over K^2}
\int_0^1 dx
{1\over R^2}= {(1-2z)\over K^2}{-1\over
\sqrt{\Delta^{(ij)}}}\ln{-2 Q_i\cdot Q_j+ \sqrt{\Delta^{(ij)}}\over
-2 Q_i\cdot Q_j- \sqrt{\Delta^{(ij)}}}. \eean
Therefore the box coefficient is given by
\bea C_{box;ij} & = & {K^2\over (1-2z)
}\left({F_{i,j}(P_{1}^{(ij)})+ F_{i,j}(P_{2}^{(ij)})\over
2}\right),~~~\label{Box-coeff}\eea
where $F_{i,j}$ is given by (\ref{box-result}).

It is important to realize that in fact (\ref{Box-coeff}) will be a rational
function of $u$ rather than a polynomial, because it contains
results for pentagons as well. To separate the box coefficient, we need to write
\bea C_{box; ij}(u)= H(u)+\sum_{i\in {\rm pentagons}} A_i P_i,~~~\label{BP-sep} \eea
where $H(u)$ is a polynomial in $u$, $P_i$ is the pentagon cut given by
(\ref{Pentagon-gen}) and $A_i$ is constant in $u$. Thus $H(u)$ will
be the true box coefficient, for which we can apply the recursion
and reduction formulas, while $A_i$ is the true pentagon coefficient.

The second term in (\ref{qqpoles}) is the final piece needed for
triangles.  When we rewrite it as
\bean & & -{1\over 2}\ln {Q_i^2\over K^2}\left.\left(
{G(\la)\prod_{s=1}^{n+k-2} \tgb{a_s|Q_i|\ell}\over \vev{\ell|K
Q_i|\ell}^{n}\prod_{t=1,t\neq i}^{k} \vev{\ell|Q_t
Q_i|\ell}}\right)\right|_{residue~of~\vev{\ell|Q_j Q_i|\ell}}  \\& &
-{1\over 2}\ln {Q_j^2\over K^2}\left.\left( {G(\la)\prod_{s=1}^{n+k-2}
\tgb{a_s|Q_j|\ell}\over \vev{\ell|K Q_j|\ell}^{n}\prod_{t=1,t\neq
j}^{k} \vev{\ell|Q_t Q_j|\ell}}\right)\right|_{residue~of~\vev{\ell|Q_i
Q_j|\ell}}, \eean
we see that
 the first term is minus half of the residue
contribution of $\vev{\ell|Q_j Q_i|\ell}$ inside  $F_i(\la)$ (
$I^{(i)}$) and the second term is minus half of the residue
contribution of $\vev{\ell|Q_i Q_j|\ell}$ inside  $F_j(\la)$
($I^{(j)}$). For the first term, when we sum up all simple pole
contributions for $F_i(\la)$, we will be left with the residue of
the possible multiple pole $\vev{\ell|K Q_i|\ell}^{n}$. That is to
say (using the fact that $Q_i^2/ K^2= x_1^{(i)}
x_2^{(i)}$),
\bea {1\over 2}\ln ( x_1^{(i)} x_2^{(i)})\left.\left(
{G(\la)\prod_{s=1}^{n+k-2} \tgb{a_s|Q_i|\ell}\over \vev{\ell|K
Q_i|\ell}^{n}\prod_{t=1,t\neq i}^{k} \vev{\ell|Q_t
Q_i|\ell}}\right)\right|_{residue~of~\vev{\ell|K
Q_i|\ell}}.~~~\label{tri-3}\eea
Summing the three triangle contributions (\ref{tri-3}) with (\ref{IKQi-2}) and
(\ref{IKQi-1-multi}) we finally reach
\bean -{1\over 2}\left.\left({G(\la)\prod_{s=1}^{n+k-2}
\tgb{a_s|Q_i|\ell}\over \vev{\ell|K Q_i|\ell}^{n}\prod_{t=1,t\neq
i}^{k} \vev{\ell|Q_t Q_i|\ell}}\ln {(-x_1^{(i)})^2\over ( x_1^{(i)}
x_2^{(i)})}\right)\right|_{residue~of~\vev{\ell~P_1^{(i)}}} \\
-{1\over 2}\left.\left({G(\la)\prod_{s=1}^{n+k-2} \tgb{a_s|Q_i|\ell}\over
\vev{\ell|K Q_i|\ell}^{n}\prod_{t=1,t\neq i}^{k} \vev{\ell|Q_t
Q_i|\ell}}\ln {(-x_2^{(i)})^2\over ( x_1^{(i)}
x_2^{(i)})}\right)\right|_{residue~of~\vev{\ell~P_2^{(i)}}}.\eean
Notice that $\ln (x_1^{(i)}/  x_2^{(i)})$ is  the
signature of triangles. To compare, we recall that for triangles we
have
\bean & & {(1-2z)}\int \vev{\ell~d\ell}[\ell~d\ell] { 1\over
\gb{\ell|K|\ell}\gb{\ell|Q_i|\ell}}= {(1-2z)}{-1\over
\sqrt{\Delta^{(i)}}}\ln{-2 Q_i\cdot K+ \sqrt{\Delta^{(i)}}\over -2
Q_i\cdot K- \sqrt{\Delta^{(i)}}}. \eean
Therefore the triangle coefficient is given by
\bea C_{tri;i} & = & {\sqrt{\Delta^{(i)}}\over
2(1-2z)}\left\{\left.\left({G(\la)\prod_{s=1}^{n+k-2}
\tgb{a_s|Q_i|\ell}\over \vev{\ell|K Q_i|\ell}^{n}\prod_{t=1,t\neq
i}^{k} \vev{\ell|Q_t
Q_i|\ell}}\right)\right|_{\vev{\ell~P_1^{(i)}}}\right. \\
& & ~~~~~~~~~~~~~~~~\left.
-\left.\left({G(\la)\prod_{s=1}^{n+k-2}
\tgb{a_s|Q_i|\ell}\over \vev{\ell|K Q_i|\ell}^{n}\prod_{t=1,t\neq
i}^{k} \vev{\ell|Q_t
Q_i|\ell}}\right)\right|_{\vev{\ell~P_2^{(i)}}}\right\}.
\eea

It is important to observe that both box and triangle coefficients
are expressed as the difference of residues at the two poles
$P_1^{(i)}$ and $P_2^{(i)}$. This  is the origin of the
``signature'' square roots of box and triangle integrals. As we will
see shortly, for bubble coefficients, we need to add up the residues at
the two poles $P_1^{(i)}$ and $P_2^{(i)}$. The addition will get rid
of the square root signature, but we are left with $(1-2z)$, which is related to bubbles.

%%%%%%%%%%%%%%%%%%%%
\subsection{Rational part}
%%%%%%%%%%%%%%%%%%%%%%%

Now we address bubble coefficients. They are given by the first
term of (\ref{Expa}) with $G_k(\la)$ given by (\ref{G-p-coe}). As we
have mentioned, for general $n$ and $k$ these are not the simplest
and most compact expressions. If we constrain $n\leq 4$, it is
possible to use them directly.  Here we use a more general approach.

Our starting point is (\ref{gen-form}).  We write it here as
\bea
\lim_{\tau \to 0}
{ G(\la) \prod_{j=1}^{n+k-2} [a_j~\ell]\over
\prod_{s=0}^{n-1}\gb{\ell|K+\tau s \eta|\ell}
\prod_{i=1}^{k}\gb{\ell|Q_i|\ell}},~~~\label{def-form}\eea
where $\eta^2=0$ and $\tau$ is a parameter which
we eventually take to zero.

Because no denominator factor appears more than once, we can use
(\ref{Split-ob-1}) directly to reach
\bea & & \sum_{s=1}^{n-1} {1\over \gb{\ell|K|\ell}\gb{\ell|K+\tau s
\eta|\ell}} {G(\la)\prod_{j=1}^{n+k-2}\tgb{a_j| K+\tau s
\eta|\ell}\over \prod_{s'=1,s'\neq s}^{n-1} \vev{\ell| (K+\tau s'
\eta)( K+\tau s \eta) |\ell}\prod_{i=1}^k \vev{\ell|Q_i ( K+\tau s
\eta)|\ell}} \nonumber \\ & & + \sum_{i=1}^{k} {1\over
\gb{\ell|K|\ell}\gb{\ell|Q_i|\ell}}
{G(\la)\prod_{j=1}^{n+k-2}\tgb{a_j| Q_i|\ell}\over
\prod_{s'=1,}^{n-1} \vev{\ell| (K+\tau s' \eta)Q_i
|\ell}\prod_{i'=1,i'\neq i}^k \vev{\ell|Q_{i'} Q_i|\ell}}.
~~~\label{rat-split}\eea
When we take the $\tau \to 0$ limit, the second line
becomes the contribution to
 triangles and boxes that was discussed in the previous
subsection. The first line gives the bubble contribution.
It appears that each term in the first line diverges in the $\tau \to 0$ limit.  To proceed, define
\bean
\W K(s)=K+\tau s \eta.
\eean
Then we find that each term in the first line of (\ref{rat-split})
can be rewritten as
\bean  {1\over \gb{\ell|\W K(s)-\tau s\eta|\ell}\gb{\ell|\W K(s)|\ell}}
{G(\la)\prod_{j=1}^{n+k-2}\tgb{a_j|\W K(s)|\ell}\over \tau^{n-2}
\vev{\ell|\eta \W K(s)|\ell}^{n-2} \prod_{s'=1,s'\neq s}^{n-1}
 (s'-s) \prod_{i=1}^k
\vev{\ell|Q_i \W K(s)|\ell}}.\eean
Now we can expand this expression as a power series in $\tau$, keeping those terms that will survive the limit $\tau \to 0$.  Specifically, we substitute
\bean {1\over \gb{\ell|\W K(s)-\tau s\eta|\ell}\gb{\ell|\W K(s)|\ell}} & = & \sum_{h=0}^{n-2}{ \tau^h s^h \gb{\ell|\eta|\ell}^h\over \gb{\ell|\W K(s)|\ell}^{2+h}}+
{\cal O}(\tau^{n-1}).
\eean
Apply the familiar spinor integration procedure to replace one integrand by a total derivative using the identity
\bea [\ell~d\ell] \left( { [\eta~\ell]^n \over
\gb{\ell|P|\ell}^{n+2}}\right)= [d\ell~\partial_{\ell}]\left(
{1\over (n+1)} {1\over \gb{\ell|P|\eta}} {[\eta~\ell]^{n+1}\over
\gb{\ell|P|\ell}^{n+1}}\right).~~~\label{Special}\eea
The result is then the
sum of residues of
\bea {G(\la)\prod_{j=1}^{n+k-2}\tgb{a_j|\W K(s)|\ell}\over \tau^{n-2}
\vev{\ell|\eta \W K(s)|\ell}^{n-1} \prod_{s'=1,s'\neq s}^{n-1}
 (s'-s) \prod_{i=1}^k
\vev{\ell|Q_i \W K(s)|\ell}}\left( \sum_{h=0}^{n-2}{ \tau^h s^h
\gb{\ell|\eta|\ell}^{h+1}\over (h+1)\gb{\ell|\W
K(s)|\ell}^{1+h}}\right).\eea
For the multiple pole,  $\vev{\ell|\eta \W
K(s)|\ell}=\vev{\ell~\eta}\tgb{\eta|\W K(s)|\ell}$.  Since
the numerator contains a factor of $[\eta~\ell]^{h+1}$, we do not
take the residue of
$\vev{\ell~\eta}$.

Finally, after defining the quantity
\bea & & R[\W K(s),\{ Q_i\}, \eta] =  \sum_{poles~except~\eta}
{\rm Res}\left({G(\la)\prod_{j=1}^{n+k-2}\tgb{a_j|\W K(s)|\ell}\over \tau^{n-2}
\vev{\ell|\eta \W K(s)|\ell}^{n-1} \prod_{s'=1,s'\neq s}^{n-1}
 (s'-s) \prod_{i=1}^k
\vev{\ell|Q_i \W K(s)|\ell}}
\right. \nonumber \\ & & ~~~~~~~~~~~~~~~~~~~~~~~~~~~~~~~~~~~~~~~~~~~~~~~~~~~~
\left.
\times
\left( \sum_{h=0}^{n-2}{ \tau^h s^h
\gb{\ell|\eta|\ell}^{h+1}\over (h+1)\gb{\ell|\W
K(s)|\ell}^{1+h}}\right)\right),\eea
we can write the bubble coefficients as
\bea C_{bubble}={1\over \sqrt{1-u}}\left. \sum_{s=1}^{n-1}R[\W
K(s),\{ Q_i\}, \eta] \right|_{\tau \to 0}. \eea
The limit is taken by expanding and truncating the series.

%%%%%%%%%%%%%%%%%%%%%%%%
\subsection{Summary of results}
%%%%%%%%%%%%%%%%%%%%%%%%

Conventions:

We start from an integrand of the form
\bea I_{term}={ G(\la) \prod_{j=1}^{n+k-2} [a_j~\ell]\over
\gb{\ell|K|\ell}^n
\prod_{i=1}^{k}\gb{\ell|Q_i|\ell}}
\eea
and present here an expression to be integrated over the final $(-2\eps)$ dimensions as discussed in Section 2.3.

Box coefficients:

\bea C_{box;ij} & = & {K^2\over \sqrt{1-u}
}\left({F_{i,j}(P_{1}^{(ij)})+ F_{i,j}(P_{2}^{(ij)})\over 2}\right)
~~~ \eea
\bea P^{(ij)}_{1,2} & = & Q_j+ y^{(ij)}_{1,2} Q_i~~~~~~~(i<j)
\eea
\bea y_{1,2}^{(ij)}& = &{-2 Q_i\cdot Q_j\pm
\sqrt{\Delta^{(ij)}}\over 2
Q_i^2},~~~~~~\Delta^{(ij)}=(2 Q_i\cdot Q_j)^2-4 Q_i^2 Q_j^2
\eea
\bea
F_{i,j}(\ell) = {G(\la)\prod_{s=1}^{n+k-2} \tgb{a_s|Q_i|\ell}\over
\vev{\ell|K Q_i|\ell}^{n}\prod_{t=1,t\neq i,j}^{k} \vev{\ell|Q_t
Q_i|\ell}}
\eea

Triangle coefficients:

\bea C_{tri;i} & = & {\sqrt{\Delta^{(i)}}\over
2\sqrt{1-u}}\left\{\left.\left({G(\la)\prod_{s=1}^{n+k-2}
\tgb{a_s|Q_i|\ell}\over \vev{\ell|K Q_i|\ell}^{n}\prod_{t=1,t\neq
i}^{k} \vev{\ell|Q_t
Q_i|\ell}}\right)\right|_{\vev{\ell~P_1^{(i)}}}\right. \nonumber \\
& & ~~~~~~~~~~~~~~~~\left. -\left.\left({G(\la)\prod_{s=1}^{n+k-2}
\tgb{a_s|Q_i|\ell}\over \vev{\ell|K Q_i|\ell}^{n}\prod_{t=1,t\neq
i}^{k} \vev{\ell|Q_t
Q_i|\ell}}\right)\right|_{\vev{\ell~P_2^{(i)}}}\right\}~~~\label{Tri-coeff}\eea
\bea
P_{1,2}^{(i)}=Q_i+ x_{1,2}^{(i)} K
\eea
\bea x_{1,2}^{(i)}& = &{-2 Q_i\cdot K\pm \sqrt{\Delta^{(i)}}\over 2
K^2},~~~~~~\Delta^{(i)}=(2 Q_i\cdot K)^2-4 Q_i^2 K^2
\eea

Bubble coefficients:

\bea C_{bubble}={1\over \sqrt{1-u}}\left. \sum_{s=1}^{n-1}R[\W
K(s),\{ Q_i\}, \eta] \right|_{\tau \to 0} ~~~~\label{Bub-coeff} \eea
The limit is taken by expanding and truncating the series.
\bea & & R[\W K(s),\{ Q_i\}, \eta] =  \sum_{poles~except~\eta}
{\rm Res}\left({G(\la)\prod_{j=1}^{n+k-2}\tgb{a_j|\W K(s)|\ell}\over \tau^{n-2}
\vev{\ell|\eta \W K(s)|\ell}^{n-1} \prod_{s'=1,s'\neq s}^{n-1}
 (s'-s) \prod_{i=1}^k
\vev{\ell|Q_i \W K(s)|\ell}}
\right. \nonumber \\ & & ~~~~~~~~~~~~~~~~~~~~~~~~~~~~~~~~~~~~~~~~~~~~~~~~~~~~
\left.
\times
\left( \sum_{h=0}^{n-2}{ \tau^h s^h
\gb{\ell|\eta|\ell}^{h+1}\over (h+1)\gb{\ell|\W
K(s)|\ell}^{1+h}}\right)\right)\eea
\bea
\W K(s)=K+\tau s \eta
\eea

For a detailed description of evaluation, see the appendix.

%%%%%%%%%%%%%%%%%%%%%%%%%%%%%%
%%%%%%%%%%%%%%%%%%%%%%%%%%%%%%%

%%%%%%%%%%%%%%%%%%%%%%%%%%%
\section{Discussion}
%%%%%%%%%%%%%%%%%%%%%%%%%%%

In this section, we first comment on the input needed for this integration program, and then discuss  the alternative approaches of generalized unitarity cuts and the program of Ossola, Papadopoulos and Pittau \cite{Ossola:2006us}.

%%%%%%%%%%%%%%%%%%%%%%%%%
\subsection{Tree level input}
%%%%%%%%%%%%%%%%%%%%%%%%%%%

The input needed for the unitary cut method is a collection of (on-shell) tree-level
amplitudes, which can be calculated by recursion relations. However,
results with  spurious poles (which generally arise in the most compact expressions) can make it hard to
apply our method. To avoid this difficulty, we can apply spinor identities to regroup the terms into an expression free of spurious poles.

Let us demonstrate this in one example. The five point function with
two massive scalars are given by\footnote{The sign of the second
term is different from the corresponding formula in
\cite{Badger:2005zh}.  We have checked, using the method of
\cite{Badger:2005zh}, that the sign in this earlier formula was a typo.}
\bean A(\ell_1, 1^+,2^+, 3^-,\ell_2) & = & -
{\gb{3|\ell_2(1+2)\ell_1|1}^2\over \vev{1~2}\vev{2~3}
((\ell_1+k_1)^2-\mu^2) ((\ell_2+k_3)^2-\mu^2) [3|(1+2)\ell_1|1]}
\\ & & + {\mu^2 [1~2]^3 \over K_{123}^2 [2~3][3|(1+2)\ell_1|1]}
\\
&=&
 -{\gb{3|\ell_2(1+2)\ell_1|1}^2\over \vev{1~2}\vev{2~3}
\gb{1|\ell_1|1}\gb{3|\ell_2|3} [3|(1+2)\ell_1|1]}
\\ & & + {\mu^2 [1~2]^3 \over K_{123}^2 [2~3][3|(1+2)\ell_1|1]}.
\eean
The spurious pole in the first term can be split from the others by the identity
\bean
{\gb{3|\ell_2(1+2)|\ell_1|1}\over
\gb{1|\ell_1|1}[3|(1+2)|\ell_1|1]}
={\vev{3|\ell_2(1+2)|1}\over \tgb{3|(1+2)|1}
\gb{1|\ell_1|1}}
+{\gb{3|\ell_2(1+2)(1+2)|3}\over \gb{1|(1+2)|3}
[3|(1+2)|\ell_1|1]}.
\eean

Then
\bean A(\ell_1, 1^+,2^+, 3^-,\ell_2)& = & - {\gb{3|\ell_2(1+2)\ell_1|1}\gb{3|\ell_2|2}\over \vev{1~2}\vev{2~3}[3~2]
\gb{3|\ell_2|3}  \gb{1|\ell_1|1}}+
{\gb{3|\ell_2(1+2)\ell_1|1}[1~2]\over \vev{1~2}\vev{2~3}[2~3]
 [3|(1+2)\ell_1|1]}\\ & & + {\mu^2 [1~2]^3 \over K_{123}^2
 [2~3][3|(1+2)\ell_1|1]}.\eean
Using
\bean \gb{3|\ell_2(1+2)\ell_1|1}
K_{123}^2
&=& \gb{3|\ell_2(1+2)\ell_1|1} (K_{12}^2-\gb{3|(1+2)|3})
\\ & = & \gb{3|\ell_2(1+2)\ell_1|1}
K_{12}^2+\gb{3|(1+2)\ell_2|3}[3|(1+2)\ell_1|1]+K_{12}^2\gb{3|\ell_1|1}
\gb{3|\ell_2|3},\eean
we can add the last two terms together and see that the spurious pole
$[3|(1+2)\ell_1|1]$ has been canceled (as it must).  Thus we finally get
\bea A(\ell_1, 1^+,2^+, 3^-,\ell_2) & = &
{\gb{3|\ell_2(1+2)\ell_1|1}\gb{3|\ell_2|2}\over
\vev{1~2}K_{23}^2\gb{1|\ell_1|1}\gb{3|\ell_2|3}}
+{[1~2]\vev{3|(1+2)\ell_2|3}\over
\vev{1~2}K_{23}^2 K_{123}^2}.~~~\label{A5-2} \eea
A similar calculation gives
\bea A_5(\ell_1^+,1^+,2^-,3^+,\ell_2^-) & = &
{\gb{2|\ell_1|1}\gb{2|\ell_2|3}^2\over \vev{1~2} K_{23}^2
\gb{3|\ell_2|3}\gb{1|\ell_1|1}}+{[3~1]\gb{2|\ell_1|1}\gb{2|\ell_2|3}\over
K_{12}^2 K_{23}^2 \gb{3|\ell_2|3}}-{\vev{2|\ell_2
K_{123}|2}[1~3]^2\over K_{12}^2 K_{23}^2
K_{123}^2},~~~\label{A5-3}\eea
where spurious poles have been canceled.

%%%%%%%%%%%%%%%%%%%%%%%%%%
\subsection{The quadruple cut}
%%%%%%%%%%%%%%%%%%%%%%%%%%

In four dimensions, the quadruple cut \cite{Britto:2004nc} is a powerful tool for getting box coefficients because the integral is completely localized.  Quadruple cuts may be applied here as well since we have separated the $(-2\eps)$-dimensional and four-dimensional integrations.  We apply them to boxes and pentagons.

Again, we write in the language of spinor integration (see the brief remarks at the beginning of Section 3), but this is absolutely not essential to the technique of quadruple cuts.

Let us start with the quadruple cut (denoted here by $C^q$) of a box. It is given by
\bean  C^q_4& = & \int_0^1 du~u^{-1-\eps} \int d^4\ell
~\delta^+(\ell^2)\delta((1-2z)K^2-2 \ell \cdot K+M_1^2-M_2^2) \\
& & \delta( (\W \ell-P_1)^2-M_3^2-\mu^2)\delta((\W
\ell-P_2)^2-M_4^2-\mu^2) \\ & = &  \int_0^1 du~u^{-1-\eps} \int
d^4\ell
~\delta^+(\ell^2)\delta((1-2z)K^2-2 \ell \cdot K+M_1^2-M_2^2) \\
& & \delta( P_1^2+M_1^2-M_3^2-z(2K\cdot P_1)-2 \ell\cdot P_1)\delta(
P_2^2+M_1^2-M_4^2-z(2K\cdot P_2)-2 \ell\cdot P_2). \eean
After $t$-integration, which sets $t=-( (1-2z) K^2+M_1^2-M_2^2) /
\gb{\ell|K|\ell}$, we have
\bean C^q_4& =& \int_0^1 du~u^{-1-\eps}\int
\vev{\ell~d\ell}[\ell~d\ell] { (1-2z) K^2+M_1^2-M_2^2 \over
\gb{\ell|K|\ell}^2}\delta\left({K^2\gb{\ell|Q_1|\ell}\over
\gb{\ell|K|\ell}}\right) \delta\left({K^2\gb{\ell|Q_2|\ell}\over
\gb{\ell|K|\ell}}\right),\eean
where
\bea
 Q_i & = & -\left( (1-2z)+{M_1^2-M_2^2\over
K^2}\right)P_1+{P_1^2+M_1^2-M_{i+2}^2-z(2K\cdot P_1)\over K^2}K .
~~~~\label{quadcutqi}
\eea
To solve the remaining two delta-functions, we need to find two
momenta constructed from $q_i=Q_2+x_i Q_1$, $i=1,2$ such that
$q_i^2=0$. This is the same construction as in (\ref{nullforqq}).
Then the two solutions for the two delta-functions are given
by\footnote{These are the $\ell_3,\ell_4$ of \cite{Ossola:2006us}.}
\bea \ell_1 = \ket{q_1} |q_2],~~~~~\ell_2 = \ket{q_2} |q_1].\eea
Notice that these solutions are complex, as usual for quadruple cuts.
There is a universal Jacobian factor, which in general
is a function of $u$. Putting it all together we have
\bea C^q_4& =& \int_0^1 du~u^{-1-\eps} J(u),~~~~\label{4-cut-box}\eea
where $J(u)$ is the Jacobian and should be symmetric in $K,P_1,P_2$.

Next we consider the quadruple cut of pentagons.
The calculation is similar, so
we shall be brief. After $t$-integration we have
\bean C^q_5& =& \int_0^1 du~u^{-1-\eps}\int
\vev{\ell~d\ell}[\ell~d\ell] { (1-2z) K^2+M_1^2-M_2^2 \over
\gb{\ell|K|\ell} K^2\gb{\ell|Q_3|\ell}}\delta\left({K^2\gb{\ell|Q_1|\ell}\over
\gb{\ell|K|\ell}}\right) \delta\left({K^2\gb{\ell|Q_2|\ell}\over
\gb{\ell|K|\ell}}\right),
\eean
with the $Q_i$ defined in (\ref{quadcutqi}).

Now we need to sum up the contributions of the two solutions. The Jacobian of
the two solutions is the same.  We end up with\footnote{Here we have
inserted the factor $1/2$ since we need to sum the contributions
of two solutions. This is the same expression used in original
quadruple cut \cite{Britto:2004nc}, where on one side, we sum up two
solutions and divide by two, while on the other side, we have just the
Jacobi factor for the basis of box.}
\bea C^q_5& =& \int_0^1 du~u^{-1-\eps} {J(u)\over K^2} {1\over
2}\left( {\gb{q_1|K|q_2}\over \gb{q_1|Q_3|q_2}}+{\gb{q_2|K|q_1}\over
\gb{q_2|Q_3|q_1}}\right)=\int_0^1 du~u^{-1-\eps} {J(u)\over 2K^2}
S[Q_2,Q_1,Q_3, K],~~~~\label{5-cut-box}\eea
where $S[Q_2,Q_1,Q_3, K]$ was defined in (\ref{Func-S}.)
We see again that the relative factor between boxes and pentagons is
 $S[\bullet]/(2K^2)$,
just as we learned from the
double cut.

In general, we perform the $t$-integral first and then add the two solutions. This gives a rational function $R(u)$.
Then we need to split
\bea R(u)= P(u)+ \sum_{Q_3} a_{Q_3} {S[Q_2,Q_1,Q_3, K]\over 2K^2},
\eea
where $P(u)$ is a polynomial that gives the box contribution, and $a_{Q_3}$ are constants in $u$ that give corresponding
coefficients of pentagons. This decomposition is the same as the one given by
(\ref{BP-sep}).

After getting box and pentagon coefficients from quadruple cuts, one might
continue by applying triple cuts to target specific triangle coefficients,
and then finally use the usual double cut for the bubble part only.
Triple cuts have been used to get
 particular one-loop coefficients in \cite{Bern:1997sc,Bern:2004ky,Bern:2004bt,Bidder:2005ri},
 and recently a nice paper \cite{Mastrolia:2006ki} has described a
 general procedure to compute triple cuts in arbitrary dimensions.
 One can try to systematically study the triple cuts of
 \cite{Mastrolia:2006ki} along the lines presented in this paper. It
 is easy to see that the delta-function there (plus possible derivatives of
 delta-function) corresponds to our multiple pole.

%%%%%%%%%%%%%%%%%%%%%
\subsection{Comparison with OPP method}
%%%%%%%%%%%%%%%%%%%%%%

Recently, a computation
by Ossola, Papadopoulos and Pittau (OPP) \cite{Ossola:2006us} has been
attracting
attention. It seems to be a very simple reduction method, which can
be performed at the integrand level. The key point of the OPP
method is that knowing the general form of spurious terms, one can
solve algebraically for the coefficients of physical and spurious terms
from
knowing the initial data.
The work we have presented here is in the same spirit.
In our method, it is through splitting into partial fractions that we are able to identify contributions
to the various basis integrals and find the functions for these coefficients.

We remark that in \cite{Ossola:2006us}, it is claimed that in the renormalizable gauge, there is an
upper bound on the number of spurious terms. For example, there are six for
triangles, eight for bubbles. The upper bound in the OPP method will
correspond to the upper bound of $n$ in our factor
$\gb{\ell|K|\ell}^n$.
Our experience suggests that we will always find
 $n\leq 4$. The reason is the following. Before the $t$-integration
we have
\bean  {\prod_{i=1}^{N_1} \tgb{a_i|\W\ell|b_i}\over
\prod_{j=1}^{N_2} ((\W \ell-P_j)^2-\mu^2)},\eean
where the power of $\W \ell$ in the numerator $N_1$ has an upper bound,
$N_1\leq N_2+2$. The reason is that in renormalizable gauge, the
expression derived from Feynman rules has the property that the degree
of $\W \ell$ in numerator is less than or equal to that in the denominator.
After peeling off the two cut propagators, we have $N_1\leq N_2+2$. Now
we count the powers of $1/\gb{\ell|K|\ell}$ after $t$-integration.
Numerators  contribute ${-N_1}$ while
denominators contribute ${N_2}$. From $\int t~
dt$ and the delta-function, we get an additional ${-2}$. In all, we have
$(1/\gb{\ell|K|\ell})^{-N_1-2+N_2}$, thus $n=N_1+2-N_2\leq 4$.

Assuming that indeed $n\leq 4$, the correspondence between our method and the OPP method becomes clearer. It is easy to see that box and
pentagon coefficients have the same expressions. For triangle, we have
multiple poles up to order $4$. From our discussion in the Appendix, it
can be seen that a pole of order $n$ can be traded for a
simple pole with an $(n-1)$-th derivative. In our case, with $n\leq 4$
we  have up to third derivatives. This is similar to
$j=1,2,3$ in \cite{Ossola:2006us}. For bubble coefficients, we have up to three terms, and
each one has multiple poles up to order three, so we would expect
$3\times 2=6$ contributions, compared to the eight spurious terms of
the OPP method.  Our impression is that the spurious terms of OPP
are
hidden as multiple poles as well as simple poles in our language. The
algebraic complexity of the two methods should be equivalent.

%%%%%%%%%%%%%%%%%%%%%%%%%%%%%%%
\acknowledgments
%%%%%%%%%%%%%%%%%%%%%%%%%%%%%%%

We wish to thank C. Anastasiou for helpful discussions and feedback on
the manuscript and R. Schabinger for informing us of several typos in
earlier versions.  RB is supported by Stichting FOM.
BF is supported by the Marie-Curie Research
Training Network under
contract MRTN-CT-2004-005104.

%%%%%%%%%%%%%%%%%%%%%%%%%
\appendix
%%%%%%%%%%%%%%%%%%%%%%

%%%%%%%%%%%%%%%%%%%%%%%%%
\section{Evaluation of residues}
%%%%%%%%%%%%%%%%%%%%%%%%%

In this appendix, we discuss how to evaluate the various functions we
defined in section 4 to get formulas for coefficients of the basis integrals.  Our aim is to make the formulas programmable, so we go into some detail.

%%%%%%%%%%%%%%%%%%%%%%%%%
\subsection{Box coefficients}
%%%%%%%%%%%%%%%%%%%%%%%%%%

The box (and pentagon) coefficients are given by (\ref{Box-coeff}),
\bea C_{box;ij} & = & {K^2\over (1-2z)
}\left({F_{i,j}(P_{1}^{(ij)})+ F_{i,j}(P_{2}^{(ij)})\over
2}\right),
\eea
where
\bean F_{i,j}(P_{1,2}^{(ij)})= {G(P_{1,2}^{(ij)})\prod_{s=1}^{n+k-2}
[a_s~P_{2,1}^{(ij)}]\over
\gb{P_{1,2}^{(ij)}|K|P_{2,1}^{(ij)}}^{n}\prod_{t=1,t\neq j,i}^{k}
\gb{P_{1,2}^{(ij)}|Q_t|P_{2,1}^{(ij)}}}.\eean
Let us rewrite the coefficient as
\bea C_{box;ij} & = & {K^2\over 2(1-2z)
}{N \over D},
~~~\label{appbox}\eea
where\footnote{Our functions $|P_{1}^{(ij)}\rangle|P_{2}^{(ij)}]$
and $|P_{2}^{(ij)}\rangle|P_{1}^{(ij)}]$ showing up in $N$ and $D$
are the  $\ell_3,\ell_4$ of \cite{Ossola:2006us}.}
\bean {N\over D} & = & F_{i,j}(P_{1}^{(ij)})+F_{i,j}(P_{2}^{(ij)}),
\\ N & = &  G(P_{1}^{(ij)})\prod_{s=1}^{n+k-2}
[a_s~P_{2}^{(ij)}]\gb{P_{2}^{(ij)}|K|P_{1}^{(ij)}}^{n}\prod_{t=1,t\neq
j,i}^{k} \gb{P_{2}^{(ij)}|Q_t|P_{1}^{(ij)}}\\ & & +
G(P_{2}^{(ij)})\prod_{s=1}^{n+k-2}
[a_s~P_{1}^{(ij)}]\gb{P_{1}^{(ij)}|K|P_{2}^{(ij)}}^{n}\prod_{t=1,t\neq
j,i}^{k} \gb{P_{1}^{(ij)}|Q_t|P_{2}^{(ij)}}, \\ D & = &
\gb{P_{1}^{(ij)}|K|P_{2}^{(ij)}}^{n}\prod_{t=1,t\neq j,i}^{k}
\gb{P_{1}^{(ij)}|Q_t|P_{2}^{(ij)}}\gb{P_{2}^{(ij)}|K|P_{1}^{(ij)}}^{n}\prod_{t=1,t\neq
j,i}^{k} \gb{P_{2}^{(ij)}|Q_t|P_{1}^{(ij)}}.\eean
For $D$ we use the following identity:
\bea \gb{\eta_1|S|\eta_2}\gb{\eta_2|S|\eta_1}= (2\eta_1\cdot
S)(2\eta_2\cdot S)-S^2(2\eta_1\cdot \eta_2).\eea
Then,
\bean
\gb{P_{1}^{(ij)}|K|P_{2}^{(ij)}}\gb{P_{2}^{(ij)}|K|P_{1}^{(ij)}} & =&
 (2K \cdot (Q_j+y_1^{(ij)} Q_i))(2K \cdot (Q_j+y_1^{(ij)} Q_i))
+{K^2 \Delta^{(ij)}\over Q_i^2}\\  & \equiv & T[Q_i,Q_j,K].\eean
The function  $T[Q_i,Q_j,K]$ in this formula is
$ T_2/Q_i^2$, with the definition of $T_2$ given in (\ref{T1-T2}).
Using this definition we have
\bea D & = & T[Q_i,Q_j,K]^n \prod_{t=1,t\neq j,i}^{k}
T[Q_i,Q_j,Q_t].\eea
It is easy to see that each term in $N$ may be written as a product of the form
\bean [a_s~P_2]\gb{P_2|Q|P_1}\vev{P_1~b_s} & = & \tgb{a_s|P_2 Q
P_1|b_s}= \tgb{a_s| (Q_j+y_2 Q_i) Q (Q_j+ y_1 Q_i)|b_s}. \eean
Thus we see that the evaluation of box coefficients (\ref{appbox}) is straightforward.

%%%%%%%%%%%%%%%%%%%
\subsection{The residue of multiple poles}
%%%%%%%%%%%%%%%%%%%

For triangles and bubbles, we need to know how to get residue of
multiple poles. There are several ways to do this. The first one is to
use the splitting method discussed in \cite{Britto:2006sj}.

The second method is  to shift momentum parametrically and then
take the limit where the parameter goes to zero.
We illustrate this method with the example of a double pole.  We denote the small parameter by $\tau$.
\bean {1\over \vev{\ell~\eta}^2} {\prod_{j=1}^k \vev{\ell~a_j}\over
\prod_{j=1}^k \vev{\ell~b_j}} & \to & {1\over
\vev{\ell~\eta}\vev{\ell~(\eta+\tau \a)}} {\prod_{j=1}^k
\vev{\ell~a_j}\over \prod_{j=1}^k \vev{\ell~b_j}} \\
& = & {1\over \vev{\eta~(\eta+\tau \a)}} {\prod_{j=1}^k
\vev{\eta~a_j}\over \prod_{j=1}^k \vev{\eta~b_j}} -{1\over
\vev{\eta~(\eta+\tau \a)}} {\prod_{j=1}^k \vev{(\eta+\tau\a)~a_j}\over
\prod_{j=1}^k \vev{(\eta+\tau\a)~b_j}} \\ &= & {1\over \tau\vev{\eta~ \a}}
{\prod_{j=1}^k \vev{\eta~a_j}\over \prod_{j=1}^k
\vev{\eta~b_j}}\left( 1- {\prod_{j=1}^k (1+ \tau {\vev{\a~a_j}\over
\vev{\eta~a_j}})\over \prod_{j=1}^k (1+ \tau {\vev{\a~b_j}\over
\vev{\eta~b_j}})}\right) \\
& \to & {1\over \vev{\eta~ \a}} {\prod_{j=1}^k \vev{\eta~a_j}\over
\prod_{j=1}^k \vev{\eta~b_j}}\left( -\sum_{j=1}^k {\vev{\a~a_j}\over
\vev{\eta~a_j}}+ \sum_{j=1}^k  {\vev{\a~b_j}\over
\vev{\eta~b_j}}\right)\\
& = &  {\prod_{j=1}^k \vev{\eta~a_j}\over \prod_{j=1}^k
\vev{\eta~b_j}}\left( \sum_{j=1}^k {\vev{a_j~b_j}\over
\vev{\eta~a_j}\vev{\eta~b_j}}\right).\eean
We see that this is the correct result. Compared to the first method, we have more
terms and one extra auxiliary spinor $\eta$. However, in this
method, the symmetry among $a_i$'s and $b_i$'s is explicit.

Now we can use the above idea to get the general expression for multiple
poles:
\bean {1\over \vev{\ell~\eta}^n} {\prod_{i} \vev{\ell~a_i}\over
\prod_j \vev{\ell~b_j}} & \to & {1\over
\prod_{s=0}^{n-1}\vev{\ell~(\eta+s \tau\a)}} {\prod_{i}
\vev{\ell~a_i}\over \prod_j \vev{\ell~b_j}} \\ & = &
\sum_{s_0=0}^{n-1}{1\over \prod_{s=0, s\neq s_0}^{n-1}\vev{(\eta+\tau
s_0\a)~(\eta+s \tau \a)}} {\prod_{i} \vev{(\eta+\tau s_0\a)~a_i}\over
\prod_j \vev{(\eta+\tau s_0\a)~b_j}} \\ & = & \sum_{s_0=0}^{n-1}{1\over
\tau^{n-1}\vev{\eta~\a}^{n-1} \prod_{s=0,s\neq s_0}^{n-1} (s-s_0)}
{\prod_{i}
\vev{(\eta+\tau s_0\a)~a_i}\over \prod_j \vev{(\eta+\tau s_0\a)~b_j}} \\
& = & \sum_{s_0=0}^{n-1}{1\over \tau^{n-1}\vev{\eta~\a}^{n-1}
\prod_{s=0,s\neq s_0}^{n-1} (s-s_0)} {\prod_{i}
\vev{\eta~a_i}(1+\tau s_0{\vev{\a~a_i}\over \vev{\eta~a_i}})\over
\prod_j \vev{\eta~b_j}(1+\tau s_0{\vev{\a~b_j}\over \vev{\eta~b_j}})} \\
& = & {1\over \tau^{n-1}\vev{\eta~\a}^{n-1}}{\prod_{i}
\vev{\eta~a_i}\over \prod_{j}
\vev{\eta~b_j}}\left(\sum_{s_0=0}^{n-1}{1\over
 \prod_{s=0,s\neq s_0}^{n-1} (s-s_0)}
{\prod_{i} (1+\tau s_0{\vev{\a~a_i}\over \vev{\eta~a_i}})\over \prod_j
(1+\tau s_0{\vev{\a~b_j}\over \vev{\eta~b_j}})}\right). \eean
We see that what we need to do is to  expand the expression
inside the parentheses as a series in $\tau$, keep only the
terms up to order
$\tau^{n-1}$. Although this expression has one auxiliary spinor $\eta$, the
final result does not depend on it. Without writing out intermediate steps,
we reach the following result:
\bea & & {1\over \vev{\eta~\a}^{n-1}}{\prod_{i} \vev{\eta~a_i}\over
\prod_{j} \vev{\eta~b_j}}\left(\sum_{s_0=0}^{n-1}{s_0^{n-1}\over
 \prod_{s=0,s\neq s_0}^{n-1} (s-s_0)}
\sum_{N_1+\sum_{j=1}^k m_j=n-1} \left( \prod_{j=1}^{k}
(-)^{m_j}\left({\vev{\a~b_j}\over
\vev{\eta~b_j}}\right)^{m_j}\right)\right.\nonumber \\& & \left.
\left( \sum_{1\leq i_1<i_2...<i_{N_1}\leq k+n-2}
\prod_{q=1}^{N_1}{\vev{\a~a_{i_q}}\over
\vev{\eta~a_{i_q}}}\right)\right).~~\label{Multi-shift}\eea

The third method is very similar to the second method. Upon noticing that
\bea  {d \over d \tau^{n-1}}{1\over \vev{\ell~\eta-\tau s}}= {(n-1)!
\vev{\ell~s}^{n-1}\over \vev{\ell~\eta-\tau s}^n},~~~\label{der}\eea
we can rewrite
\bean {1\over \vev{\ell~\eta}^n} {\prod_{i} \vev{\ell~a_i}\over
\prod_j \vev{\ell~b_j}} & \to & {1\over \vev{\ell~\eta-\tau s}^n}
{\prod_{i} \vev{\ell~a_i}\over \prod_j \vev{\ell~b_j}} \\
& = &  {d \over d \tau^{n-1}}\left({1\over \vev{\ell~\eta-\tau s}} {1\over
(n-1)! \vev{\ell~s}^{n-1}} {\prod_{i} \vev{\ell~a_i}\over \prod_j
\vev{\ell~b_j}}\right).\eean
Now we take the simple residue of $\vev{\ell~\eta-\tau s}$.  That is, we set
$\ket{\ell}=\ket{\eta-\tau s}$, take the derivative, and  finally
find that the residue is
\bea {\rm Residue}={d \over d \tau^{n-1}}\left( {1\over (n-1)!
\vev{\eta~s}^{n-1}} {\prod_{i} \vev{(\eta-\tau s)~a_i}\over \prod_j
\vev{(\eta-\tau s)~b_j}}\right)_{\tau\to 0}.~~~\label{Multi-der} \eea
It should not be hard to check that (\ref{Multi-shift}) is the same
as (\ref{Multi-der}) if we identify $\a=s$.
As a demonstration, let
us check the case of a double pole:
\bean & & {1\over \vev{\ell~\eta-\tau s}^2} {\prod_{j=1}^k
\vev{\ell~a_j}\over \prod_{j=1}^k \vev{\ell~b_j}}  \to {d \over d
\tau}\left.\left({1\over \vev{\ell~s}\vev{\ell~\eta-\tau s}}
{\prod_{j=1}^k \vev{\ell~a_j}\over \prod_{j=1}^k
\vev{\ell~b_j}}\right)\right|_{residue}
\\ & =  & {d \over d \tau}\left.\left({1\over
\vev{\eta-\tau s~s}}{\prod_{j=1}^k \vev{\eta-\tau s ~a_j}\over
\prod_{j=1}^k \vev{\eta-\tau  s~b_j}}\right)\right|_{\tau\to 0} \\ & = &
{1\over \vev{\eta~s}}{\prod_{j=1}^k \vev{\eta ~a_j}\over
\prod_{j=1}^k \vev{\eta~b_j}}\left( \sum_{j} {\vev{s~b_j}\over
\vev{\eta~b_j}}-\sum_j {\vev{s~a_j}\over \vev{\eta~a_j}}\right).\eean
This is exactly the same result given by the shifting technique.

Knowing how to evaluate the residue of multiple poles, we can discuss
how to evaluate coefficients for triangles and bubbles given by
(\ref{Tri-coeff}) and (\ref{Bub-coeff}). We  use the third method.
For example, for (\ref{Tri-coeff}) we can shift $K\to
K- \tau s$. Then using (\ref{der}) we can transfer it to the residue of
a simple pole. We can further  simplify the sum of these two simple
poles using the same technique discussed in our previous subsection
for box coefficients. After that we can take the derivative and
the $\tau \to 0$ limit by truncating the power series.

%%%%%%%%%%%%%%%%%%%%%%%%%%%%%%%%%%%%%%%%%%%%%%%%%%%%

\end{document}